\documentclass[iop]{emulateapj}

\usepackage{hyperref}

\usepackage{graphicx}
\usepackage{epsfig}
\usepackage{natbib}
\usepackage{multirow}
\usepackage{amsmath,amssymb}
\usepackage{rotating}
\usepackage{pdfpages}
\usepackage{xcolor}
\usepackage{url}

\begin{document}

\submitted{Accepted to The Astrophysical Journal}
\journalinfo{Draft version \today}

\title{An ALMA Survey of Submillimeter Galaxies in the Extended Chandra Deep Field South: Near-infrared morphologies and stellar sizes}
\author{Chian-Chou Chen\altaffilmark{1}, Ian Smail\altaffilmark{1}, A. M. Swinbank\altaffilmark{1}, J. M. Simpson\altaffilmark{1}, Cheng-Jiun Ma\altaffilmark{1}, D. M. Alexander\altaffilmark{1}, A. D. Biggs\altaffilmark{2}, W. N. Brandt\altaffilmark{3,4}, S. C. Chapman\altaffilmark{5}, K. E. K. Coppin\altaffilmark{6}, A. L. R. Danielson\altaffilmark{1}, H. Dannerbauer\altaffilmark{7}, A. C. Edge\altaffilmark{1}, T. R. Greve\altaffilmark{8}, R. J. Ivison\altaffilmark{2,9}, A. Karim\altaffilmark{10}, Karl M. Menten\altaffilmark{11}, E. Schinnerer\altaffilmark{12}, F. Walter\altaffilmark{12}, J. L. Wardlow\altaffilmark{13}, A. Wei{\ss}\altaffilmark{11}, P. P. van der Werf\altaffilmark{14}}

\altaffiltext{1}{Institute for Computational Cosmology, Durham University, South Road, Durham DH1 3LE UK.}
\altaffiltext{2}{European Southern Observatory, Karl Schwarzschild Strasse 2, D-85748 Garching, Germany.}
\altaffiltext{3}{Department of Astronomy \& Astrophysics, 525 Davey Lab, Pennsylvania State University, University Park, PA 16802, USA.}
\altaffiltext{4}{Institute for Gravitation and the Cosmos, Pennsylvania State University, University Park, PA 16802, USA.}
\altaffiltext{5}{Department of Physics and Atmospheric Science, Dalhousie University, Halifax, NS B3H 3J5, Canada.}
\altaffiltext{6}{Centre for Astrophysics Research, Science and Technology Research Institute, University of Hertfordshire, Hatfield AL10 9AB, UK.}
\altaffiltext{7}{Universit\"{a}t Wien, Institut f\"{u}r Astrophysik, T\"{u}rkenschanzstra{\ss}e 17, A-1180 Wien, Austria.}
\altaffiltext{8}{Department of Physics and Astronomy, University College London, Gower Street, London WC1E 6BT, UK.}
\altaffiltext{9}{Institute for Astronomy, University of Edinburgh, Royal Observatory, Blackford Hill, Edinburgh EH9 3HJ, UK.}
\altaffiltext{10}{Argelander-Institute for Astronomy, Bonn University, Auf dem H\"{u}gel 71, D-53121 Bonn, Germany.}
\altaffiltext{11}{Max-Planck-Institut f\"{u}r Radioastronomie, Auf dem H\"{u}gel 69, D-53121 Bonn, Germany}
\altaffiltext{12}{Max-Planck Institute for Astronomy, K\"{o}nigstuhl 17, D-69117 Heidelberg, Germany.}
\altaffiltext{13}{Dark Cosmology Centre, Niels Bohr Institute, University of Copenhagen, Denmark.}
\altaffiltext{14}{Leiden Observatory, Leiden University, P.O. Box 9513, 2300 RA Leiden, The Netherlands}
\subjectheadings{cosmology: observations|  galaxies: formation  |  galaxies: starburst  | submillimeter: galaxies }

\begin{abstract}
We analyse {\it HST} WFC3/$H_{160}$-band observations of a sample of 48 ALMA-detected submillimeter galaxies (SMGs) in the Extended Chandra Deep Field South field, to study their stellar morphologies and sizes. We detect 79\,$\pm$\,17\% of the SMGs in the $H_{160}$-band imaging with a median sensitivity of 27.8 mag, and most (80\%) of the non-detections are SMGs with 870\,$\mu$m fluxes of $S_{870} < $\,3\,mJy. With a surface brightness limit of $\mu_H \sim $\,26\,mag\,arcsec$^{-2}$, we find that 82\,$\pm$\,9\% of the $H_{160}$-band detected SMGs at $z = $\,1--3 appear to have disturbed morphologies, meaning they are visually classified as either irregulars or interacting systems, or both. By determining a S\'{e}rsic fit to the $H_{160}$ surface-brightness profiles we derive a median S\'{e}rsic index of $n = $\,1.2\,$\pm$\,0.3 and a median half-light radius of $r_e = $\,4.4$^{+1.1}_{-0.5}$\,kpc for our SMGs at $z = $\,1--3. We also find significant displacements between the positions of the $H_{160}$-component and 870\,$\mu$m emission in these systems, suggesting that the dusty star-burst regions and less-obscured stellar distribution are not co-located. We find significant differences in the sizes and the S\'{e}rsic index between our $z = $\,2--3 SMGs and $z \sim $\,2 quiescent galaxies, suggesting a major transformation of the stellar light profile is needed in the quenching processes if SMGs are progenitors of the red-and-dead $z\sim$\,2 galaxies. Given the short-lived nature of SMGs, we postulate that the majority of the  $z = $\,2--3 SMGs with $S_{870} \gtrsim $\,2\,mJy are early/mid-stage major mergers.

\end{abstract}

\section{Introduction}
Bright submillimeter galaxies (SMGs) with $S_{870} \gtrsim$\,2\,mJy represent a population of distant, dust-obscured galaxies which were most prevalent around 10\,Gyr ago ($z\sim$\,2; e.g., \citealt{Smail:1997p6820, Barger:1998p13566, Hughes:1998p9666, Chapman:2005p5778, Wardlow:2011qy, Yun:2012aa,Simpson:2014aa}). Through extensive, multi-wavelength observations, it appears that SMGs have many characteristics similar to local ultra-luminous infrared galaxies (ULIRGs), such as their  gas fractions, far-infrared luminosities and  interstellar medium (ISM) properties \citep{Tacconi:2008p9334, Danielson:2011aa, Bothwell:2013lp, Riechers:2011aa, Casey:2014aa}. With implied star-formation rates (SFRs) of $\sim$ 500\,M$_{\odot}$\,yr$^{-1}$ SMGs have the potential to form half of the stars in a massive galaxy (M$_{\star}>$\,10$^{11}$\,M$_{\odot}$) in just $\sim$\,100\,Myr.  Moreover, over the redshift range $z$\,=\,1--4, bright SMGs contribute $\sim$\,10--30\% of the total star-formation budget (e.g., \citealt{Barger:2012lr, Casey:2013aa, Swinbank:2014ul}). Inevitably, links have therefore been made between SMGs and passive elliptical galaxies (e.g., \citealt{Lilly:1999lr, Fu:2013mm, Simpson:2014aa}). SMGs have also been linked to AGN/QSO activity at $z\sim$\,2 (e.g., \citealt{Alexander:2005p6453, Coppin:2008pp, Simpson:2012aa, Wang:2013aa}) and the formation of the compact quiescent galaxy population being found at $z = $\,1--2 (e.g., \citealt{Cimatti:2008aa, Whitaker:2012ab, Toft:2014aa}).  

However, there are also important differences between SMGs and local ULIRGs. In particular, the spatial extent of the gas and star-formation in SMGs appears to be much larger than that typically seen in local ULIRGs ($\sim$\,few kpc in SMGs compared to just 100's of parsecs in local ULIRGs; e.g., {\citealt{Charmandaris:2002aa, Sakamoto:2008aa, Menendez-Delmestre:2009aa, Diaz-Santos:2010aa, Engel:2010p9470, Ivison:2011aa, Rujopakarn:2011aa}}). This could be due to the star formation occurring either in extended disks in high-redshift, pre-coalesce mergers, or in tidal features associated with a late-stage merger. One route to testing the link between local ULIRGs and high-redshift SMGs is by comparison to galaxy formation models. However, theoretical models attempting to reproduce the basic properties of bright SMGs (such as the number counts and redshift distribution) have come to a variety of conclusions regarding the physical processes that trigger the star formation: low-mass merging starbursts, with unusually low mass-to-light ratios \citep{Baugh:2005p14519, Swinbank:2008fj}; isolated (or not-strongly interacting), gas-rich disk galaxies with secular bursts (e.g., \citealt{Dave:2010kx, Cowley:2014aa}); and a hybrid scenario where some SMGs are merger-induced starbursts and some secularly evolving disks (e.g., \citealt{Hayward:2011fr, Hayward:2013qy}). 

To distinguish chaotic systems such as mergers from ordered rotating disks, in principle we can compare the rotational velocity and velocity dispersions measured through emission lines originating from cool molecular gas and/or ionised gas, as well as searching for irregular morphologies, tidal features, or spiral arms in the stellar light in deep rest-frame optical/near-infrared  imaging. Many dynamical studies using emission lines from the $^{12}$CO molecule or H$\alpha$ have concluded that they see evidence of mergers (e.g., \citealt{Tacconi:2008p9334, Engel:2010p9470, Alaghband-Zadeh:2012aa, Menendez-Delmestre:2013aa}). However, individual cases of clumpy, rotating disks have also been found (e.g., \citealt{Swinbank:2011aa, Hodge:2012fk}), although these do not preclude a merger origin (e.g., {\citealt{Robertson:2008aa, Ueda:2014aa}}). Moreover, the drawback of studying kinematics based on molecular or ionised gas is that these tracers are sensitive to the dynamics of the interstellar medium (ISM), in which the dense molecular gas could appear as rotating disks even in merging systems (e.g., NGC 3256, \citealt{Sakamoto:2006aa}). Moreover, the dynamics of the ionised gas could be heavily affected by strong outflows (e.g., \citealt{Harrison:2012aa}). A complementary approach is therefore to map the stellar distribution, the collisionless tracer of the galaxy morphology, to investigate the formation mechanism of bright SMGs. 

Previous studies of the stellar morphologies of SMGs using high-resolution imaging from the {\it Hubble Space Telescope} ({\it HST}) have told a mixed story. The earliest optical {\it HST} studies of SMGs were confused by the counterpart-identification problems which have plagued SMG samples (e.g. \citealt{Smail:1998aa}). Using radio-located SMG samples with more robust counterpart-identifications, \citet{Chapman:2003aa} and \citet{Conselice:2003aa} studied the morphologies of SMGs using rest-frame UV imaging.   However, the rest-frame UV suffers strong  obscuration in these very dusty sources, biasing the conclusions. 

A key advance came with high-resolution near-infrared imaging from {\it HST} which provides a less dust sensitive probe of the stellar distribution on 0.1$''$ (sub-kpc) scales. Such analyses with NICMOS have shown that SMGs have apparently compact and disturbed morphologies, with evidence for tidal tails and highly asymmetric light distributions \citep{Swinbank:2010aa, Conselice:2011aa, Aguirre:2013aa}. {However, at $z > $\,1 star-forming galaxies are often disturbed (e.g., \citealt{Bluck:2012aa}), thus such traits do not necessarily indicate an on-going merger as would be the case locally (e.g., \citealt{Mortlock:2013aa})}. Unfortunately, the faintness of SMG counterparts combined with the relatively shallow surface brightness limits of the NICMOS imaging (1-$\sigma$ $\mu_H\sim$\,24.5\,mag\,arcsec$^{-2}$), meant that these studies could not differentiate the properties of the SMGs from those of relatively quiescent UV-selected galaxies at similar redshifts \citep{Giavalisco:2002aa, Lotz:2008aa, Law:2012aa}.

More recent studies of the stellar morphology of SMGs have used deeper $H_{160}$-band imaging from the new Wide Field Camera 3 (WFC3) camera on {\it HST} with a substantially deeper surface brightness sensitivity,  $\mu_H \sim$\,26 mag arcsec$^{-2}$. Various quantitative analyses have been conducted, including concentration ($C$), asymmetry ($A$), clumpiness ($S$) (CAS) method \citep{Conselice:2003ab}, the Gini/M20 parameters, as well as fitting the light profile to determine the S\'{e}rsic indices and effective radius. {In these studies, SMGs are found to have an average S\'{e}rsic index of $n \sim $\,1, analogous to the local disky galaxies \citep{Targett:2013aa}. \citet{Wiklind:2014aa} also argue} that SMGs represent a more isolated, asymmetric, and heterogeneous population in contrast with those dusty sources at similar redshifts but selected at 100/160\,$\mu$m \citep{Kartaltepe:2012aa}.

{Unfortunately even some of these recent morphological studies rely on samples of candidate SMGs selected} from coarse resolution ($\sim$\,20$''$) of single-dish submillimeter surveys,   which are  identified through
 indirect, empirical correlations between the submillimeter emission and that at other wavelengths where higher angular resolution observations are possible, such as the radio or mid-infrared (e.g. \citealt{Ivison:1998p10286, Smail:2000p6377, Ivison:2002uq, Pope:2005uq, Biggs:2011uq}). While the success rate of these indirect techniques is upto 80\,\%, they  suffer both contamination from false identifications and  various selection biases  (e.g., \citealt{Wang:2011p9293, Hodge:2013lr}).

%
%
\begin{figure*}
\begin{center}
    \leavevmode
      \includegraphics[scale=0.45]{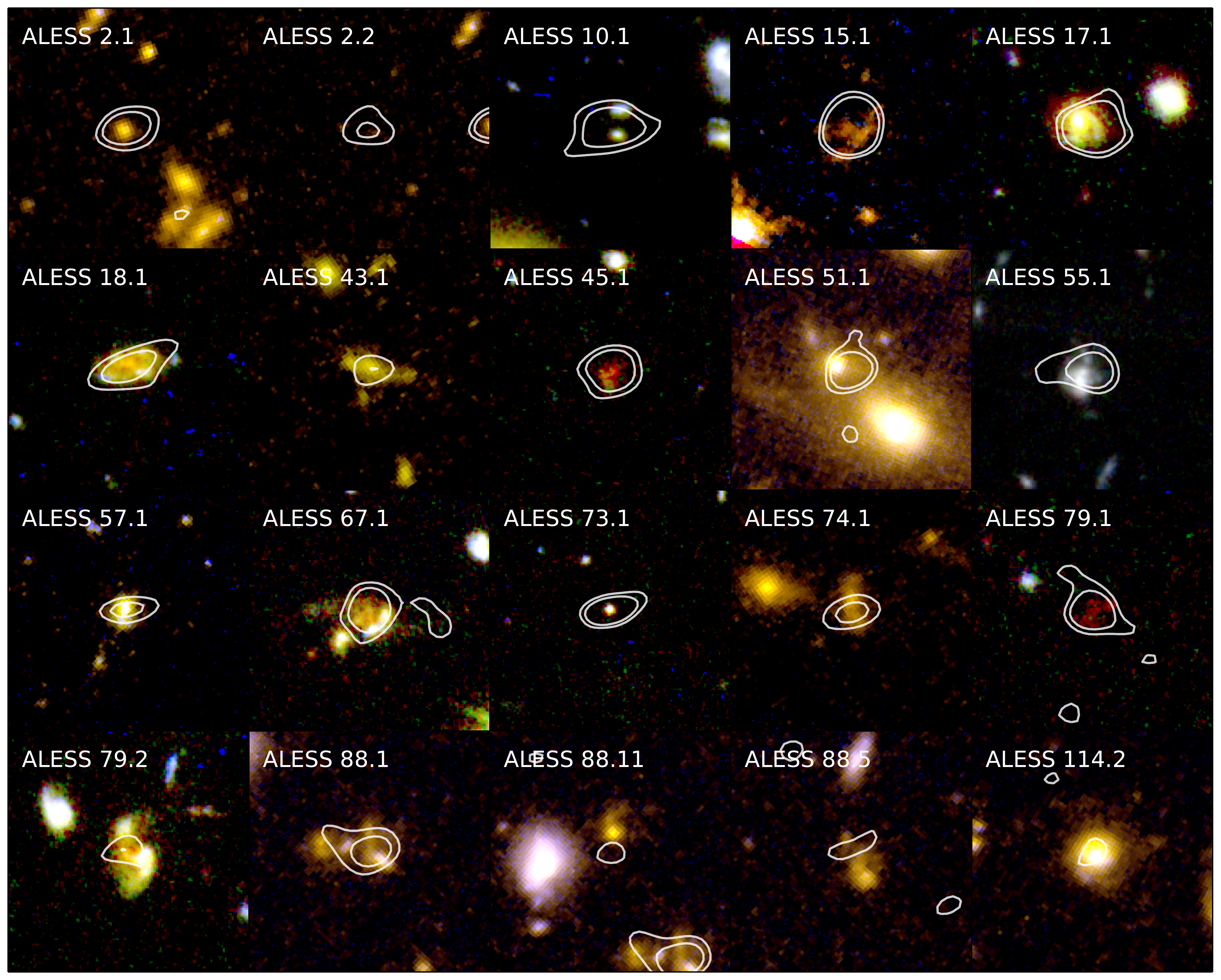}      
       \caption{{\it HST} false-color images of {20 out of 25 $H_{160}$-detected ALESS SMGs that have imaging  in at least two {\it HST}  bands, to demonstrate the diversity in morphologies present within the sample.} In these maps the red and blue channels are  WFC3 $H_{160}$ and ACS $I_{814}$, respectively, except for ALESS\,55.1 where the blue is WFC3 $J_{105}$. The green channel is WFC3 $J_{125}$ if available, or if not we made artificial green maps fby interpolating  $H_{160}$ and  $I_{814}$. These images show that SMGs have a mix of morphology classes; but the majority, $\sim$\,80\%, appear to be irregulars or interacting systems in the rest-frame optical. The contours show the submillimeter emission from our ALMA 870-$\mu$m maps with levels at 3 and 5\,$\sigma$. The size of each box is $\sim$\,70\,$\times$\,70\,kpc based on the photometric redshifts ($\sim $\,8$''\times$\,8$''$ at $z \sim $\,2).}
    \label{rgb}
 \end{center}
\end{figure*}

The only secure method for precisely locating the SMGs from a single-dish continuum survey is interferometric observations in the same waveband as the original discovery survey. The first interferometric studies of SMGs in the sub/millimeter were undertaken soon after their discovery, but they are observationally expensive and were initially limited to small samples \citep{Frayer:2000aa, Gear:2000aa, Lutz:2001aa, Dannerbauer:2002aa, Dannerbauer:2008aa, Iono:2006p6985, Wang:2007p6971, Wang:2011p9293, Younger:2007p6982, Younger:2008rt, Younger:2009p9502, Cowie:2009p6978, Aravena:2010p8370, Hatsukade:2010aa, Chen:2011p11605, Chen:2014aa, Barger:2012lr, Smolcic:2012pp, Smolcic:2012lr}. In addition, some interferometric millimeter line surveys have also been undertaken both to locate the gas reservoir associated with the strongly star-forming SMGs and to simultaneously confirm their redshifts (e.g., \citealt{Frayer:1998fj, Greve:2005p6788, Bothwell:2013lp}), but these have typically targeted emission in low-$J$ $^{12}$CO transitions and hence may not unambiguously identify the precise location of the dust continuum counterparts (e.g., \citealt{Tacconi:2006p8449, Ivison:2010yq}). 

We have undertaken an ALMA Cycle 0 band 7 (870\,$\mu$m) study of the first large sample of submillimeter sources, a flux-limited sample of 126 submillimeter sources in the $0.5^{\circ} \times 0.5^{\circ}$ Extended {\it Chandra} Deep Field South (ECDFS), taken from the {LABOCA submillimeter single-dish survey at 870\,$\mu$m in ECDFS (``LESS'' survey; \citealt{Weis:2009qy})}. These ALMA maps directly pin-point the position of each SMG to sub-arcsecond accuracy, free from the uncertainties due to the use of probabilistic radio/mid-infrared associations \citep{Karim:2013fk, Hodge:2013lr}. To study the stellar morphologies we have collected a set of deep {\it HST} WFC3 $H_{160}$ band imaging on 48 of our ALESS SMGs with 1-$\sigma$ depths of $\mu_H\sim$\,26\,mag\,arcsec$^{-2}$, $\sim$\,3 times deeper than the surface brightness limits of  previous NICMOS studies. 
The unambiguous identifications from the ALMA observations of a uniformly selected sample of SMGs, in combination with the improved near-infrared imaging capability of WFC3 on board {\it HST} now put us in a position to readdress the question of the morphological properties of SMGs and to test the theoretical models which predict different triggering mechanisms.

%
%
\begin{figure*}
\begin{center}
    \leavevmode
      \includegraphics[scale=1]{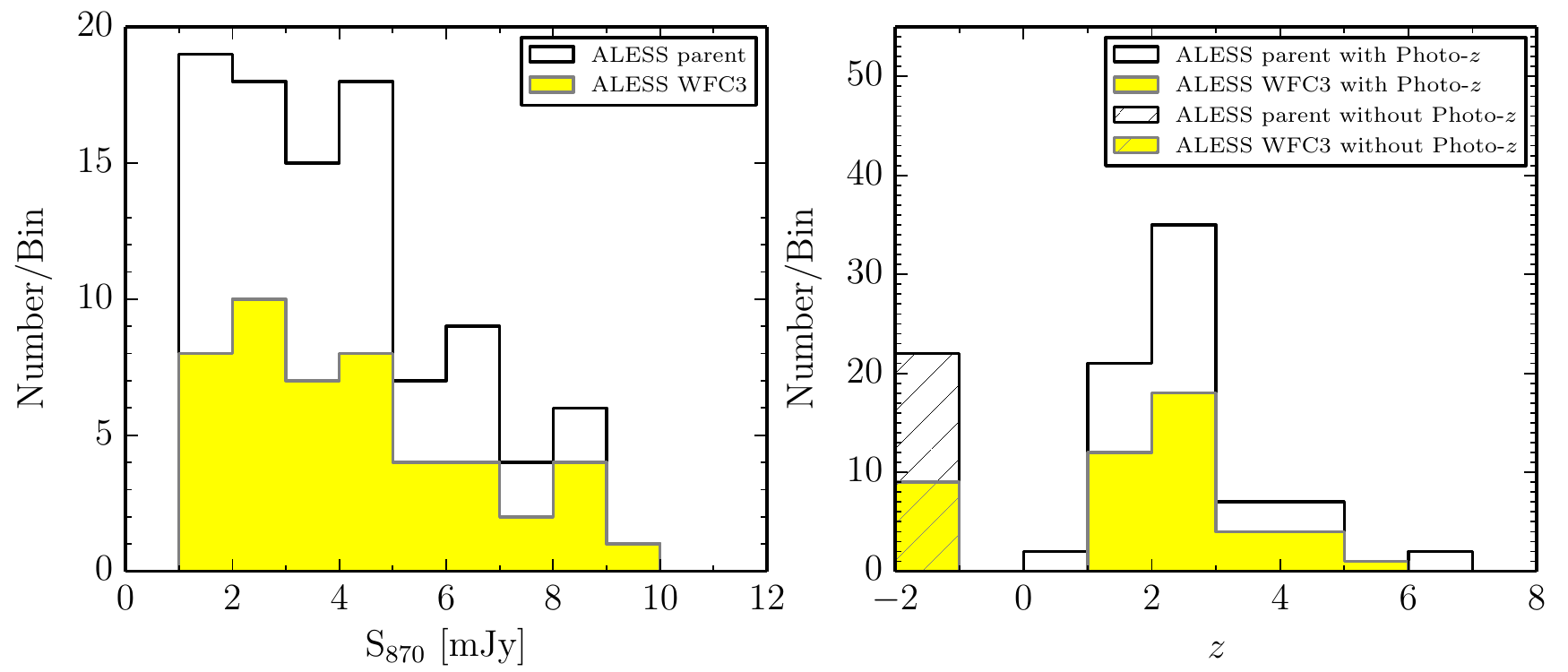}
       \caption{{\it Left}: Histogram showing the distribution of fluxes from the MAIN catalog of the ALESS survey \citep{Hodge:2013lr} (black solid line). The yellow region shows the distribution of fluxes for the MAIN ALESS SMGs which were covered by our {\it HST} WFC3/$H_{160}$-band imaging. {\it Right}: Photometric redshift distribution for the MAIN ALESS SMGs (black solid line). The median redshift for the 77 MAIN ALESS SMGs with photometric redshift measurements is $z = $\,2.3\,$\pm$\,0.1, with a tail out to $z > $\,5 \citep{Simpson:2014aa}. The yellow region again shows the redshift distribution for the {\it HST} WFC3/$H_{160}$-band observed MAIN ALESS SMGs. The hatched region shows those MAIN ALESS SMGs that do not have redshift measurements (manually assigned to $z = -$2). The K-S test shows the {\it HST}-observed ALESS SMGs are drawn from the parent ALESS SMG population with a confidence level of $>$\,99.5\%. Note we treated the two pairs ALESS\,55.1/55.5 and ALESS\,67.1/67.2 as two single sources in this paper (see \autoref{sec:hst} for details).}
    \label{sample}
 \end{center}
\end{figure*}

We describe the data and our methodology in \autoref{sec:ooda}. The results are given in \autoref{sec:results}. We discuss the implications of our results in \autoref{sec:dis} and summarize this paper in \autoref{sec:sum}. Throughout this paper we adopt the AB magnitude system \citep{Oke:1983aa}, and we assume the {\it Wilkinson Microwave Anisotropy Probe} cosmology: H$_0 = $\,70.5\,km\,s$^{-1}$\,Mpc$^{-1}$, $\Omega_M =$\,0.27, and $\Omega_\Lambda =$\,0.73 \citep{Larson:2011ys}.

\section{Sample, Observations, and Data Reduction}\label{sec:ooda}
Our sample is drawn from the ALMA study of the LESS sources  (ALESS) which provide the first large-scale, unbiased identifications for a complete sample of SMGs. The full ALESS sample was presented in \citet{Hodge:2013lr}, which defined a robust ``MAIN'' catalog of 99 SMGs that are: 1) located within the primary beam of ALMA; 2) detected with S/N\,$\geq$\,3.5; 3) detected in the ALMA maps with a 1-$\sigma$ r.m.s.\ of less than 0.6\,mJy\,beam$^{-1}$; and 4) have a major-to-minor axis ratio of the synthesized beam of $\leq $\,2. In addition, they also provided a ``SUPPLEMENTARY'' catalog of 32 SMGs that meet part of the criteria for the MAIN sample (i.e., S/N\,$\geq$\,4 and have an r.m.s.\ $>$\,0.6\,mJy/beam, lie outside the primary beam, or have a major-to-minor axis ratio $>$ 2).

\subsection{{\it Hubble Space Telescope} Observations}\label{sec:hst}
The new {\it HST}/WFC3 observations were carried out in the $H_{160}$ band during Cycle 20 from late-2012 to mid-2013 (PID: 12866; PI: A.M.\ Swinbank) under LOW-SKY conditions in all exposures to minimise the background due to zodiacal light and Earth shine in order to probe the structural properties of any low surface brightness features. There were 15 pointings and the exposure time was two orbits per pointing. The data were processed through the standard pipeline reduction using {\sc AstroDrizzle}. {We also made use of images taken as part of two {\it HST} legacy programs, the Cosmic Assembly Near-infrared Deep Extragalactic Legacy Survey (CANDELS; \citealt{Grogin:2011fj, Koekemoer:2011aa}) and the {\it Hubble} Ultra Deep Field 2009 (HUDF09) program \citep{Bouwens:2011vn}.} The images taken by both programs have either similar or deeper depths in the $H_{160}$-band compared to our new data.

To calibrate the astrometry of our {\it HST} images, we aligned the $H_{160}$-band images to the 3.6-$\mu$m image of the full ECDFS from {\it Spitzer}/IRAC, the common astrometric frame  adopted in \citet{Simpson:2014aa}. We used {\sc SExtractor} to create a source catalogue for each {\it HST} image, and the 3.6\,$\mu$m imaging. We match each source in the {\it HST} catalogue to the 3.6\,$\mu$m  catalogue, and measure an individual offset for each image. We apply a median offset of $\Delta $R.A.\,$=$\,0$\farcs$11 and $\Delta $Dec.\,$= -$0$\farcs$25, and all applied offsets are $<$\,1$''$. We statistically tested the accuracy of our astrometry by calculating the offsets between randomly chosen 3.6\,$\mu$m sources and their $H_{160}$-band counterparts lying within a 0$\farcs$8 radius circle (equivalent to the IRAC/3.6\,$\mu$m PSF), and we found no systematic offset and a scatter of $\sim$\,0$\farcs$16 in both R.A.\ and Dec., consistent with the expected accuracy of the IRAC imaging \citep{Damen:2011lr}.

{We used {\sc SExtractor} \citep{Bertin:1996zr} to extract sources from our {\it HST} imaging, and set the detection threshold to be 16 connecting 1\,$\sigma$ pixels. We adopted the automatic aperture magnitudes ({\sc MAG\_AUTO}) for the flux measurements. Our {\sc SExtractor} settings on the detection threshold have yielded a range of $H_{160}$-band sensitivity between 27--30 mag with a median sensitivity of $\sim$\,27.8\,mag. The sensitivity was estimated by summing the background r.m.s.\ values assigned to each pixel by {\sc SExtractor}, which are the r.m.s.\ values {\sc SExtractor} used to trigger detections, within the 4$\times$4 pixel box (our detection threshold of 16 connecting 1\,$\sigma$ pixels) centred at the ALMA position.}

\subsection{Our sample}\label{sec:sample}
To make efficient use of telescope time, we selected pointings for which there are two (or more) SMGs covered by the WFC3 field of view. We do not expect this to bias our results: as the photometric (and spectroscopic) redshifts of the SMGs in each WFC3 field do not suggest that the sources are physically associated. With 15 pointings, we have covered 48 ALESS SMGs. When combined with the ten ALESS SMGs covered by CANDELS and the two ALESS SMGs observed by HUDF09 this brings the total sample to 60, in which 48 (12) sources are from the  MAIN (SUPPLEMENTARY) catalog. In this study, our conclusions are drawn from the 48 ALESS SMGs in the MAIN catalog, a reliable catalog with $\sim$\,99\% completeness and a false detection rate of $\sim$\,1.6\% \citep{Karim:2013fk}.  In addition we also present the results for the ALESS SMGs in the  SUPPLEMENTARY catalog in the Appendix. {Note that two other ALESS SMGs were also covered by our {\it HST} observations, ALESS\,55.5 and ALESS\,67.2. However in our $H_{160}$-band imaging both appear to be physical associations of the same SMG; ALESS\,55.5 is an extended clump eastward of ALESS\,55.1, and ALESS\,67.2 appears to be a separated detection $\sim$2$''$ westward of ALESS\,67.1.} We therefore treated the two pairs, ALESS\,55.1/55.5 and ALESS\,67.1/67.2, as single sources in our analysis.

In Figure~\ref{rgb} we show example thumbnails of 20 ALESS SMGs covered by CANDELS/HUDF09. Our sample size is at least a factor of  two larger than any previous near-infrared SMG morphological studies, as well as benefiting from unambigious and unbiased interferometric submillimeter identifications \citep{Swinbank:2006aa, Swinbank:2010aa, Aguirre:2013aa, Targett:2013aa, Wiklind:2014aa}. 

The histograms of 870\,$\mu$m fluxes and redshifts of the 48 $H_{160}$-band observed ALESS SMGs are shown in Figure~\ref{sample} along with those of the ALESS parent population. A K-S test shows that the {\it HST}-observed subsample is drawn from the ALESS parent sample with a confidence level of $>$\,99.5\%.  For these tests we summed the 870\,$\mu$m fluxes of ALESS\,55.1/55.5 and ALESS\,67.1/67.2 and we adopted the redshifts of ALESS\,55.1 and ALESS\,67.1 for the two pairs.

The photometric redshifts ($z_{photo}$) of the ALESS SMGs were derived using the SED fitting code {\sc hyperz} \citep{Bolzonella:2000aa} using the observed photometry from ultraviolet (UV) to mid-infrared (MIR) wavelengths (for details see \citealt{Simpson:2014aa}). In total, 39 out of 48 $H_{160}$-band observed ALESS SMGs have well-constrained photometric redshifts. {This is confirmed by a comparison to the spectroscopic redshifts from Danielson et al.\ (in preparation), who find a median $\Delta z / (1 + z_{\rm spec}) = -0.004 \pm 0.026$}. Together with the photometric redshifts, we used the 250--870\,$\mu$m photometry from \citet{Swinbank:2014ul} to derive infrared luminosities ($L_{\rm IR}$) for those 39 $H_{160}$-band observed ALESS SMGs \citep{Swinbank:2014ul}.

\subsection{Non-SMG comparison sample}\label{sec:compsample}
To provide a control sample to compare to the SMGs, we also analysed a sample of field galaxies in an identical manner to the ALESS SMGs. We exploit the Multi-wavelength Survey by Yale-Chile (MUSYC) catalog, in which 32-band optical to mid-infrared photometric measurements are provided along with the derived photometric redshifts \citep{Cardamone:2010aa}. To define our sample, we selected any galaxies in the MUSYC catalog that are located within the ALMA primary beam but undetected in the ALMA imaging ($S_{870} \lesssim $\,2\,mJy). We only select sources that are located at $z=$\,1--3, the redshift range in which we focus our analysis. We adopted the spectroscopic redshifts if available, and for sources with only photometric redshifts we made a quality cut of $Q_z < $\,2, as suggested by the author of {\sc eazy} (the program used for deriving the photometric redshift) because the scatter increases sharply above $Q_z = $\,2 and the 5\,$\sigma$ outlier fraction on sources with $Q_z > $\,2 increases to around 30\% \citep{Brammer:2008aa}. Within these selection limits, there are 58 galaxies. We note that on average the comparison sample has a redshift distribution that is skewed to slightly lower redshift, with $<\! z \!> = $\,1.5$^{+0.9}_{-0.3}$, than the SMGs.

\section{Methodology}\label{sec:meth}
Studies of galaxy morphology often involve analysis using quantitative tools such as the CAS method \citep{Conselice:2003ab}, the Gini coefficient \citep{Abraham:2003aa}, and the M20 parameter \citep{Lotz:2004aa}. However, although quantitative methods are useful tools for building objective references and separating bulge dominated early-type galaxies from disk-like late-type galaxies, on some occasions shapes or features that are obvious to the eye, and critical for separating irregulars/mergers from disks, are hard to distinguish with these methods. For example, CAS analysis in the optical $R$-band imaging is virtually incapable of {distinguishing} between local late-type disks, irregulars, and edge-on disks \citep{Conselice:2003ab}, as the non-parametric method is only sensitive to rankings of the pixel value, not permutations of the pixel position \footnote{For example, the Gini coefficient is used to measure the distribution of light among pixels, with higher values of the coefficient indicating an unequal distribution (Gini of 1 means all light is located in one pixel), while lower values indicate more even distributions (Gini of 0 means every pixel has an even share). The Gini value is calculated using the Lorentz curve \citep{Lorentz:1905aa} of the galaxies' light distribution, which does not take any spatial information into consideration. We refer the reader to \citet{Conselice:2014aa} for a full review on non-parametric methods.}. One can imagine a face-on disk with spiral arms having similar CAS values as a face-on irregular with the same pixel value distribution of that face-on disk but with pixel positions shuffled. Indeed, at $z > $\,1 late-type galaxies are shown to be often disturbed \citep{Bluck:2012aa}, and CAS analysis can not separate disks from irregulars \citep{Mortlock:2013aa}, which could be the reason that previous morphological studies of SMGs found separating $z \sim$\,2 UV selected galaxies from $z \sim$\,2 SMGs difficult using Gini/M20 or CAS \citep{Swinbank:2010aa, Wiklind:2014aa}. 

This caveat on non-parametric analysis motivates the need to first conduct a {visual inspection of} the morphology of SMGs before we measure the size and surface brightness profiles (e.g., \citealt{van-Dokkum:2010aa}). Hence we next study our sample of ALESS SMGs using visual classifications, along with the surface brightness fitting tool, {\sc GALFIT}, to measure effective radius (half-light radius) and S\'{e}rsic indices.  

\subsection{Visual Classifications}\label{sec:vis}
%
%
\begin{figure}
\begin{center}
    \leavevmode
      \includegraphics[scale=0.37]{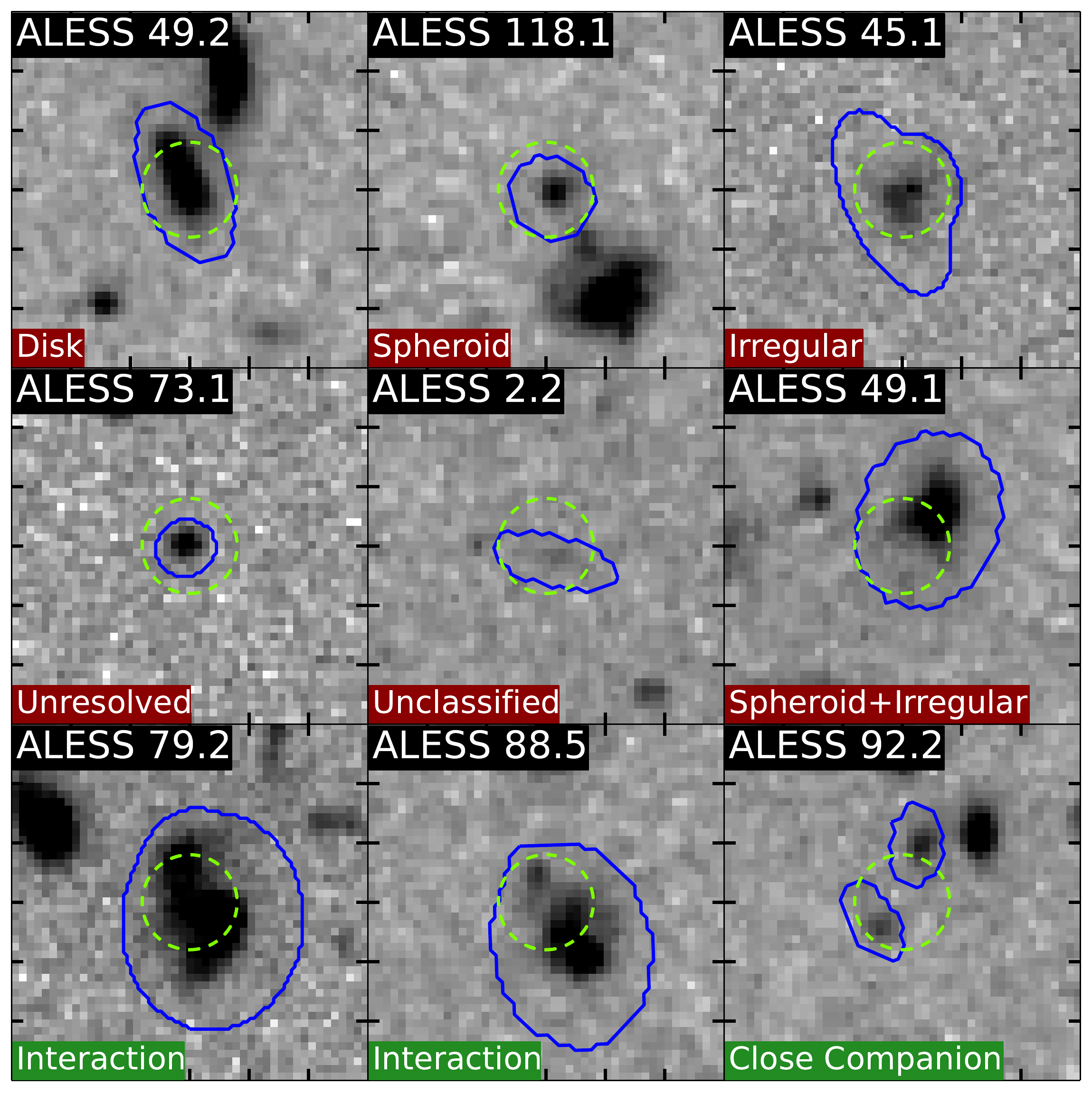}
       \caption{Example thumbnails illustrating the different morphological classifications. Each panel is $6'' \times 6''$, and the tick marks show 1$''$ increments. Green dashed circles show the typical ALMA beam (1$\farcs$6 FWHM), and the blue contours enclose the regions where we fit the S\'{e}rsic profile (see \autoref{sec:galfit} for detail). {The main morphological classes are shown in red  while the two additional classification tags are labelled in green.} Images of all 60 $H_{160}$-band observed ALESS SMGs can be found in Appendix A.}
    \label{example}
 \end{center}
\end{figure}

Following recent $H_{160}$-band morphological studies in the CANDELS fields (e.g., \citealt{Kartaltepe:2012aa, Kocevski:2012aa}), we visually classified the $H_{160}$-band morphologies of our {\it HST}-observed ALESS SMGs into five main morphological classes, which are described as follows: 

\begin{enumerate}
\item Disk -- sources with apparent disk-like structure, regardless of whether they display spiral arms, or having an {elongated morphology with or without a }central bulge. 
\item Spheroid -- sources with apparent circular/ellipsoidal structure with resolved, centrally concentrated emission.
\item Irregular -- sources with structures that do not belong to any of the other classes, or sources that are classified as disks or spheroids but have  disturbed structures in the outskirts (noted as disk+irregular or spheroid+irregular). 
\item Unresolved -- sources with structures that are well-fit by a point spread function (PSF; detailed model fits are described in \autoref{sec:galfit}). 
\item Unclassified -- sources that are contaminated by  nearby bright stars, galaxies, image artifacts  or are too faint to be classifiable. 
\end{enumerate}
Note that the first three classes are not mutually exclusive; one source can be classified as a combination of different classes, such as Disk + Spheroid or Disk + Spheroid + Irregular.

If there are multiple counterparts located within the ALMA synthesized beam of a single SMG, we tagged those sources as either ``interaction'' or ``close companions''. Sources that have tidal tails or low level emission connecting multiple counterparts were tagged as an interaction, while those that do not have apparent interacting features were tagged as close companions. {An interaction tag may also represent a source that appears to be in the final coalescent stage of a merger, with asymmetric merger remnants.} We show examples for each class along with the two additional tags in Figure \ref{example} {(see also Figure 4 and 5 in \citealt{Kartaltepe:2014aa})}. These tags are independent from the main classifications described above. For example, an SMG such as ALESS\,49.1 can be classified as Spheroid + Irregular with an interaction tag.

Four of our team members (CCC, IRS, JMS, CJM)  determined  visual classifications of the $H_{160}$-detected SMGs, as well as the comparison sample, and we derived from their classifications the median fraction of all main morphological classes as well as the bootstrapped errors, which are later used in our analysis.

\subsection{{\sc galfit}}\label{sec:galfit}
To quantify the morphology of the SMGs we fitted a S\'{e}rsic profile to the $H_{160}$-band surface brightness of each individual counterpart galaxy, using the most recent version of {\sc GALFIT} (v3.0.5, \citealt{Peng:2010aa}). {\sc galfit} is a two-dimensional (2D) fitting algorithm that is designed to fit the  surface brightness distribution of a source with various pre-defined models such as the S\'{e}rsic profile \citep{de-Vaucouleurs:1948aa, Sersic:1968aa}, which is described as:

\begin{equation}
\Sigma(r) = \Sigma_e\exp\left[-\kappa\left(\left(\frac{r}{r_e}\right)^{1/n}-1\right)\right]
\label{eqa:sersic}
\end{equation}

\noindent
where $ \Sigma_e$ is the surface brightness at a effective radius $r_e$, $n$ is the S\'{e}rsic index indicating the concentration of a light profile with higher values indicating more concentrated profiles, and $\kappa$ normalises the fit so that the effective radius ($r_e$) equals the half-light radius where half of the total source flux is emitted within $r_e$. {We report the $r_e$ measured along the semi-major axis, which is the direct output of {\sc galfit}.}

The key {\sc galfit} inputs for each source are a science image, PSF map,  ``sigma'' (or weight) map, mask outlining pixel regions to be considered in the model fitting, and the first guess for the parameters of the S\'{e}rsic profile. Below we describe how we generated these inputs in detail.

The best way to generate a PSF is to median stack nearby unsaturated and background subtracted bright stars that are Nyquist sampled ({\sc fwhm} $>$\,2 pixels; \citealt{Morishita:2014aa}). This can be achieved in CANDELS and HUDF09, as mosaics were obtained through sub-pixel dithering resulting in a finer pixel scale. We therefore generated the PSF images following this procedure for sources covered by these two datasets. However, our newly obtained {\it HST} maps have a pixel resolution of $\sim$\,0$\farcs$13. Comparing to the $H_{160}$-band PSF size ({\sc fwhm} $\sim$\,0$\farcs$18) they are not Nyquist sampled. To circumvent this issue, we first used {\sc TinyTim} \citep{Krist:2011aa}{\bf, a standard PSF simulation software package for {\it HST},} to model the PSF profiles based on the pixel locations of our sources. The PSF profiles were modelled with a pixel resolution three times higher than that of the original {\it HST} images (oversampling factor of 3). We then fitted stars close to our sources by allowing {\sc GALFIT} to convolve the model PSFs with Gaussian functions. Finally we used the best-fit Gaussian functions to convolve the model PSFs to generate the properly centred, effective PSFs that we later used for S\'{e}rsic fits.

Input science images were cut into 20$'' \times $\,20$''$ thumbnails centred at the ALMA SMG positions. The local background sky and dispersions were measured by fitting Gaussian functions to the pixel distribution of the thumbnails.

For our new WFC3 observations, we used {\sc galfit} to generate the sigma images. In order to do so {\sc galfit} requires the science images to be calibrated in units of Electrons ($e^-$). We therefore convert the unit of our drizzled images from $e^-$\,s$^{-1}$ to $e^-$ by multiplying the drizzled images by the exposure time. We generated sigma images for sources covered by CANDELS and HUDF09 by taking the inverse square root of the weight  maps. The weight maps of CANDELS/HUDF09 imaging represent pure inverse variance including all background noise terms (sky level for each exposure, read noise, dark current, and flat-field structure). However, the final combined mosaics also contain significant amounts of correlated noise, mostly due to the fact that the PSF is undersampled by the detector pixels. Smaller pixel scale mosaics were created through sub-pixel dithering, however at the cost of correlated noise. We therefore manually scaled the CANDELS/HUDF09 sigma images such that the variance of the signal-to-noise distribution was unity, and derived correction factors of typically $\sim$\,40--50\%. Note we also used the sigma maps generated by {\sc galfit} instead of the weight maps for CANDELS-observed SMGs using the simulated exposure maps publicly available on the CANDELS website, and we found no significant difference in our results.
 
The input science thumbnails were masked during fitting to prevent the background sky from dominating the $\chi^2$ values. We used {\sc SExtractor} to generate the segmentation maps used in masking.
Each detected source is described as an elliptical shape in {\sc SExtractor}, and a set of parameters, $C_{XX}$, $C_{XY}$, and $C_{YY}$, are provided in the output catalogs. The parameters are used to parametrize the elliptical shape as $C_{XX}(x-x_0)^2+C_{YY}(y-y_0)^2+C_{XY}(x-x_0)(y-y_0)=R^2$, where $R$ is a scaling parameter, in units of semi-major axis (or semi-minor axis), and $x_0$ and $y_0$ are pixel positions of the source center. The $H_{160}$-band counterparts of the ALESS SMGs are selected to be any source that is located within the radius of a typical ALMA synthesized beam (0$\farcs$8). We generated the masks based on the shape parameters of each counterpart, with $R$ ranging from 3--6 depending on the structures of each source. The masks are outlined in blue contours in Figure \ref{example}.

The initial values for the S\'{e}rsic profile are based on the flux, semi-major, semi-minor axis, and the positional angle in the output catalogs of {\sc SExtractor}. In general, for each source we first obtained a well-constrained fit with the initial parameters by varying the mask scaling parameter $R$. We settled with the smallest $R$ that allowed us to obtain a fit. We then used the parameters of this fit as the input for the next iteration. We repeated this process until the output parameters converged. Note that we normally did not place restrictions on the range of any parameters during the fit. However, due to complicated structures such as tidal tails or very disturbed/faint surface brightness in some systems, mostly involving a multi-component fit, we fixed the S\'{e}rsic index to either 1 or 4 (depending on the prefered index early in the iteration process) on a small number of components. Fixing this index prevents {\sc galfit} from producing unconstrained results with abnormally large/small values that are likely to be unphysical (e.g., a S\'{e}rsic index of 20 with an $r_e$ of 200\,kpc). As discussed in {\autoref{subsec:vis}}, fixing the S\'{e}rsic index in the model fit for a small number of sources ($\sim$ 15\%) does not affect our conclusions.

\section{Results}\label{sec:results}
We summarise the results of our visual classification and profile fitting of the 48 $H_{160}$-observed ALESS SMGs in Table~\ref{tab:galfit}. In total we detected 38 SMGs\footnote{Note ALESS\,5.1 is strongly contaminated by a nearby bright galaxy and our {\sc SExtractor} settings could not distinguish the two. We therefore measured the $H_{160}$-band flux of ALESS\,5.1 using aperture photometry with a 4-pixel ($\sim$\,0$\farcs$54) diameter aperture and a local sky estimate. We varied the sky annulus to estimate the systematic uncertainties, which were included in the errors, and selected the value that produced the best signal-to-noise ratio (S/N\,$\sim$\,6).}, with a detection rate of 79\,$\pm$\,17\%. In addition, 61\,$\pm$\,16\,\% (23 out of 38) of the $H_{160}$-detected ALESS SMGs have more than one $H_{160}$-band component based on \autoref{tab:galfit}. {In contrast}, only $\sim$\,10\% (5 out of 58) of the comparison sample have multiple $H_{160}$-band components.

%
%
\begin{figure}
\begin{center}
    \leavevmode
      \includegraphics[scale=0.4]{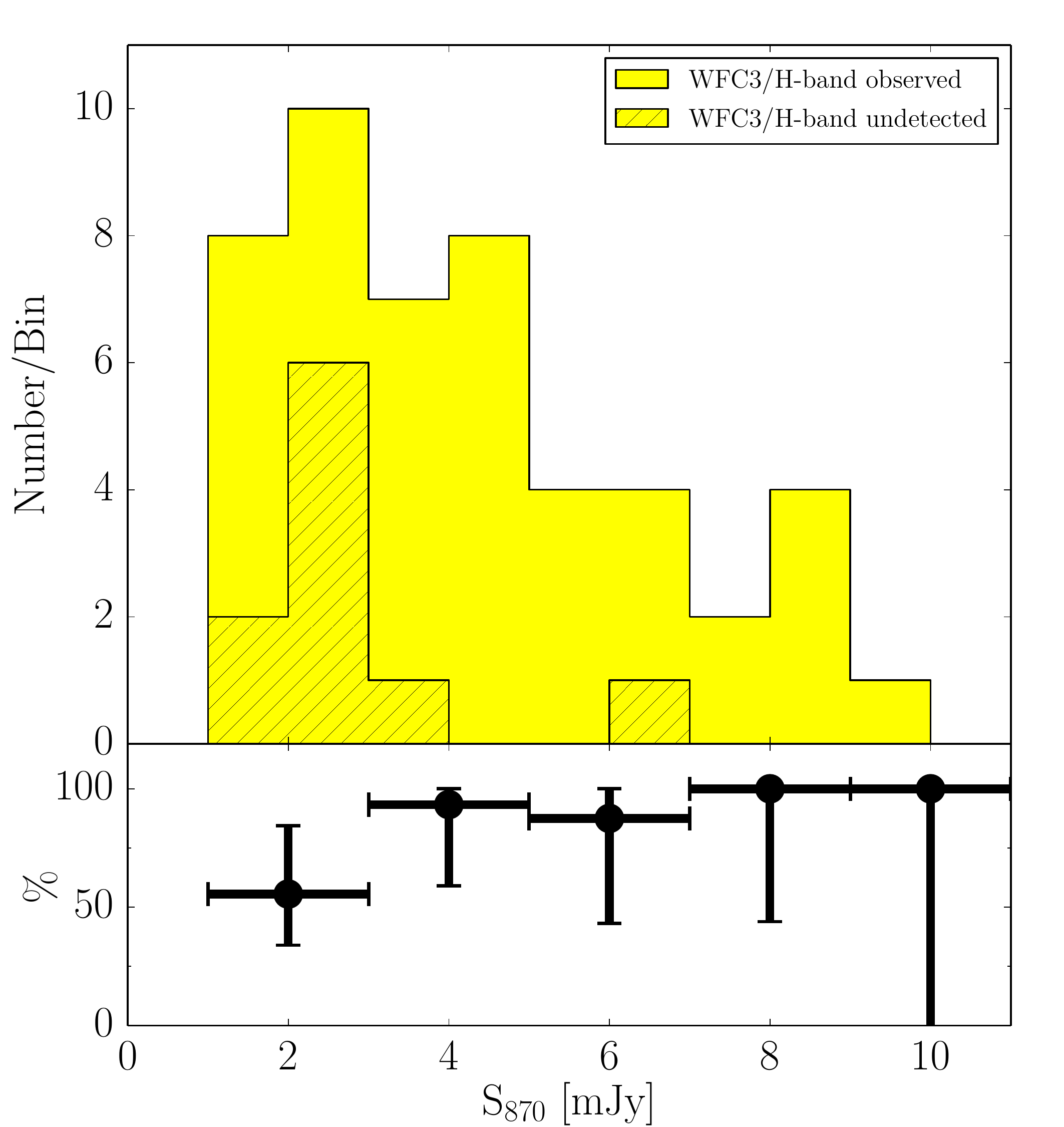}
       \caption{{\it Upper}: Histogram of the 870\,$\mu$m flux distribution of the $H_{160}$-band observed, and the $H_{160}$-band undetected, ALESS SMGs. {\it Lower}: The $H_{160}$-band detection rate of the ALESS SMGs in each flux bin. The errors were estimated through Poisson statistics. Most ($\sim$\,80\%) of the $H_{160}$-undetected SMGs are submillimeter faint with $S_{870} < $\,3\,mJy.}
    \label{detecrat}
 \end{center}
\end{figure}

%
%
\begin{figure*}
\begin{center}
    \leavevmode
      \includegraphics[scale=1.1]{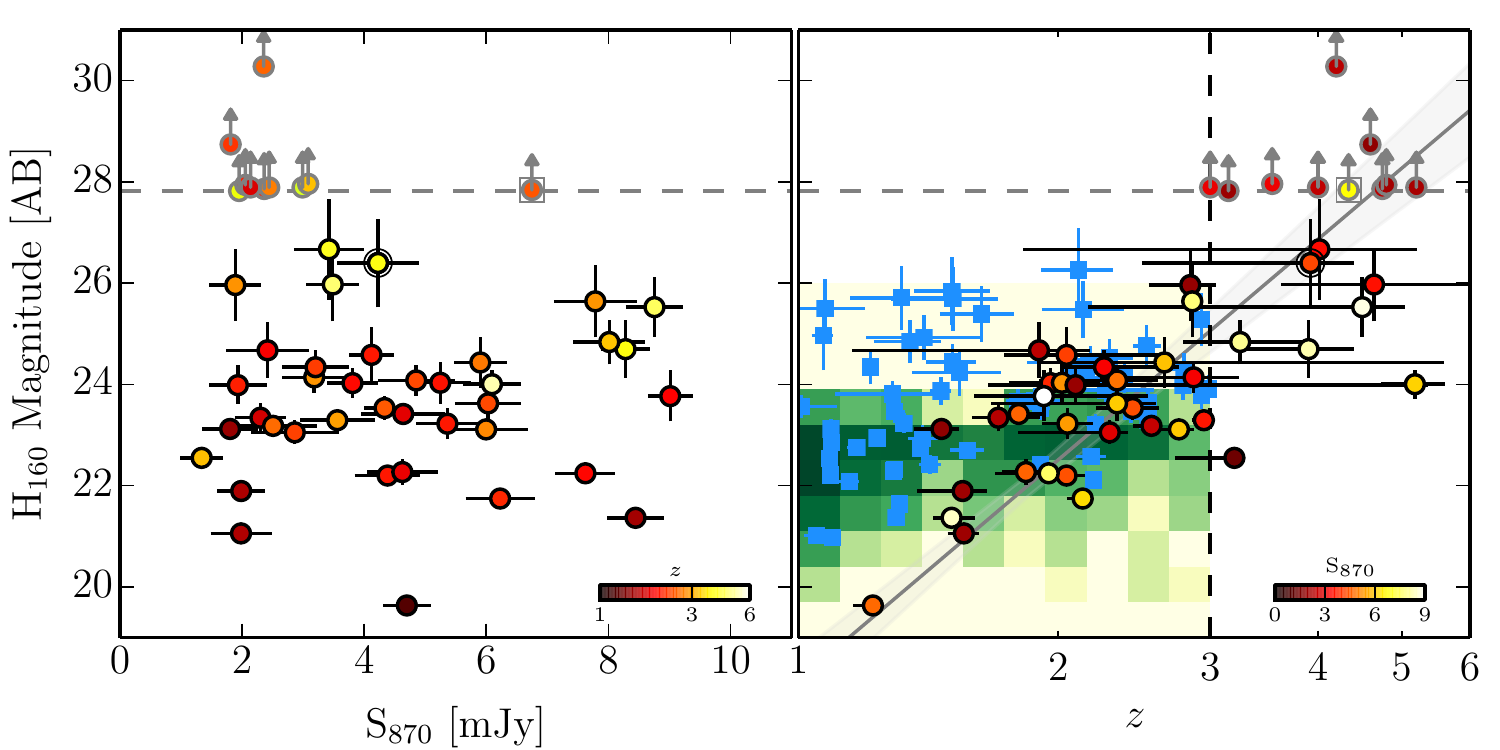}
       \caption{{\it Left}: Total $H_{160}$-band magnitude versus 870\,$\mu$m flux for the 48 $H_{160}$-band observed ALESS SMGs. The median detection threshold of the $H_{160}$-band imaging is shown as the grey dashed line (27.8), and the non-detections (10) are plotted as grey {circles with limits}. Each SMG is color coded based on their redshift. ALESS\,1.1, which is potentially lensed by nearby foreground galaxies, is additionally enclosed by a grey square. In addition, ALESS\,2.2, enclosed by a black circle, is the only source which is detected in $H_{160}$-band but has an unconstrained photometric redshift, and for this a crude redshift estimate is derived from the relation between the rest-frame $H_{160}$-band magnitude and redshift seen for ALESS SMGs at $z < $\,2.5 \citep{Simpson:2014aa}.  {\it Right}: $H_{160}$-band magnitude versus redshift, with each data point color coded based on its 870\,$\mu$m flux. While there is no obvious correlation between the $H_{160}$-band magnitude and the 870\,$\mu$m flux, a positive correlation exists between $H_{160}$-band magnitude and redshift, and the best linear regression fit is shown as a grey line with 1\,$\sigma$ errors in shading. The vertical black dashed line marks the redshift $z = $\,3 below which our $H_{160}$-band flux measurements are complete. The blue {squares} are the results of the ALESS comparison sample as described in \autoref{sec:compsample} and the 2-dimensional histogram in green shows the source number density based on the MUSYC catalog in the ECDFS field \citep{Cardamone:2010aa}. The ALESS SMGs are among the brightest $H_{160}$-band sources at $z < $\,2, {possibly} suggesting that they are among the most massive systems across this redshift range.}
    \label{hmag}
 \end{center}
\end{figure*}

In Figure \ref{detecrat}, we plot the 870\,$\mu$m flux distribution of the ALESS SMGs that were observed in $H_{160}$-band, as well as those that were undetected in our $H_{160}$-band imaging. As shown in the figure, most of the $H_{160}$ non-detections are for SMGs with $S_{870}\sim $\,1--3\,mJy, and the fraction of $H_{160}$-band undetected ALESS SMGs increases at fainter 870\,$\mu$m fluxes (Figure \ref{detecrat}). A similar trend of decreasing detection rate for fainter ALESS SMGs was also seen at the radio/mid-infrared wavelengths \citep{Hodge:2013lr}. In fact, this trend continues to even fainter SMGs with $S_{870} <$\,1\,mJy \citep{Chen:2014aa}. These $H_{160}$-undetected  SMGs are potentially high-redshift sources which are too faint to be detected in infrared/radio data at the current sensitivities \citep{Dannerbauer:2002aa, Chen:2014aa, Simpson:2014aa}. 

We plot in  Figure \ref{hmag} the $H_{160}$-band magnitude against 870\,$\mu$m flux of the ALESS SMGs, color coded by their redshifts. Although the {\sc SExtractor} and {\sc galfit} measured flux densities for all the $H_{160}$-band detected ALESS SMGs (except ALESS\,5.1 due to strong foreground contamination) are in agreement, for the total $H_{160}$-band flux densities we adopted those measured using {\sc SExtractor} since they are direct measurements and are not model dependent. Again, most of the $H_{160}$-band undetected SMGs have $S_\text{870} < $\,3\,mJy, and in fact, the brightest SMG that is not detected in the $H_{160}$-band imaging, ALESS\,1.1 (see Figure \ref{hmag}), is located only a few arcseconds away from a bright foreground galaxy, suggesting that it may be gravitationally lensed and its intrinsic flux may be  fainter. While there is no obvious correlation between $H_{160}$-band magnitude and 870\,$\mu$m flux, a trend of SMGs with fainter $H_{160}$ magnitudes lying at higher redshifts can be seen. Based on the relation between the rest-frame $H_{160}$-band magnitude and redshift for ALESS SMGs at $z < $\,2.5, \citet{Simpson:2014aa} crudely estimated the redshifts (adopted in this plot) of these $H_{160}$-band undetected ALESS SMGs and suggested that most of them are likely to be at $z > $\,3. 

We can see this trend more clearly in the right panel of Figure \ref{hmag}, where we plot $H_{160}$-band magnitude against redshift for the ALESS SMGs. Note that ALESS\,2.2 (enclosed by a circle in both panels) is the only SMG that is detected in $H_{160}$-band but unconstrained in photometric redshift, and the statistically estimated redshift described above is adopted (for the illustrative purpose only). A lack of low-redshift, $H_{160}$-faint SMGs is unlikely to be a selection effect given that our $H_{160}$-band imaging is deep enough to detect them, if they exist. On the other hand, we might miss a few very rare high-redshift $H_{160}$-band bright galaxies that a survey with a larger area coverage than our original LESS survey is capable of detecting. At $z < $\,3, where our $H_{160}$-band flux measurements are complete (Figure \ref{hmag}), we derive a non-parametric Spearman's rank correlation coefficient of 0.51 with a probability of 99.6\% ($p$-value of 0.4\%) that a correlation exists between $H_{160}$-band magnitude and redshift. We conducted a linear least-square fits to our data at $z=$\,1--3 (where the $H_{160}$-band measurements are complete) and found a best fit relation of $H = (18.2\pm 0.6)+(14.5\pm 2.0) \times \text{log}_{10}z$ with a reduced $\chi^2 = $\,0.7. A lack of correlation in the $H_{160}$--$S_{870}$ diagram implies no or weak correlation between $S_{870}$ and $z$, which confirms the finding of \citet{Simpson:2014aa}, however again with the caveat that the redshifts for the fainter SMGs are  incomplete.

The correlation we found for our ALESS SMGs between the $H_{160}$-band magnitude and  redshift is in general agreement with the previous SMG studies, showing a similar trend between $K$-band magnitude and redshift \citep{Dannerbauer:2004aa, Smail:2004aa, Clements:2004qy, Chen:2013fk, Barger:2014aa}. {Interestingly, at $z < $\,2, comparing to either the comparison sample or the field galaxies across the whole ECDFS from the MUSYC catalog, the correlation appears to trace the bright $H_{160}$-band envelop of the field populations, possibly suggesting that SMGs are among the most luminous and hence potentially massive systems at $z < $\,2. At $z > $\,2, ALESS SMGs occupy a similar locus in the $H$--$z$ diagram as the field population. However, the intrinsic $H_{160}$-band flux of $z > $\,2 SMGs could be much higher given that at these redshifts the $H_{160}$-band traces rest-frame wavelengths blueward of the $V$-band, where the SMGs will suffer more significantly from dust obscuration. Deep and longer wavelength observations are needed to investigate whether SMGs are also among the most brightest, and thus the most massive systems, at $z > $\,2.}

Our $H_{160}$-band flux measurements of the ALESS SMGs are complete up to $z = $\,3 thanks to the deep $H_{160}$-band imaging. To make a clean assessment of our sample of SMGs we now focus our analysis on the 29 (excluding ALESS\,5.1 due to strong contamination) $H_{160}$-band detected ALESS SMGs at $z=$\,1--3 (Figure \ref{hmag}), unless otherwise stated. The results for $z > $\,3 SMGs are included in \autoref{subsec:visualclass} for illustration, and our results are in fact insensitive to the redshift cut. By making this redshift cut we include most of the SMGs that are bright enough in the $H_{160}$-band for a visual classification, and we found that more than 95\% of the $H_{160}$-detected ALESS SMGs at $z < $\,3 are classifiable. In addition, at $z < $\,3 the $H_{160}$-band imaging effectively maps rest-frame optical emission ($\lambda_{rest} > $\,4000\,\AA), and hence provides a more reliable tracer of the stellar light distribution.

\subsection{Visual Classifications}\label{subsec:visualclass}

The visual classification method described in \autoref{sec:vis} shows that 
many of our systems have structural elements which fall in more than one  morphological class, reflecting the complexity of their structures (as a result, because the  classification bins are not mutually exclusive, the percentages of SMGs classed in different morphological bins in the following do not sum to 100\%).  

Our visual classification finds that 61$^{+12}_{-11}$\% of the $H_{160}$-band detected ALESS SMGs at $z = $\,1--3 include a component with a disk-like morphology {(including pure disks and any combination of disk and the other main classes such as disk+spheroid or disk+irregular). We also find that 37$^{+10}_{-8}$\% of the SMGs are either pure spheroids or include a component with spheroidal morphology, and that 58$^{+13}_{-12}$\% are pure irregulars or include a component which appears irregular.} Moreover, 52$^{+11}_{-10}$\% of the SMGs appear to have features indicative of interaction. Combining the latter two classes, we find that the percentage of SMGs with disturbed morphologies, meaning they are either irregular or have an interaction tag, is 82\,$\pm$\,9\%. {In contrast}, we found that only 48\,$\pm$\,6\% of the comparison far-infrared-faint field sample have disturbed morphologies, suggesting that these are relatively dynamically calmer or more isolated systems.

%
%
\begin{figure}
\begin{center}
    \leavevmode
      \includegraphics[scale=0.47]{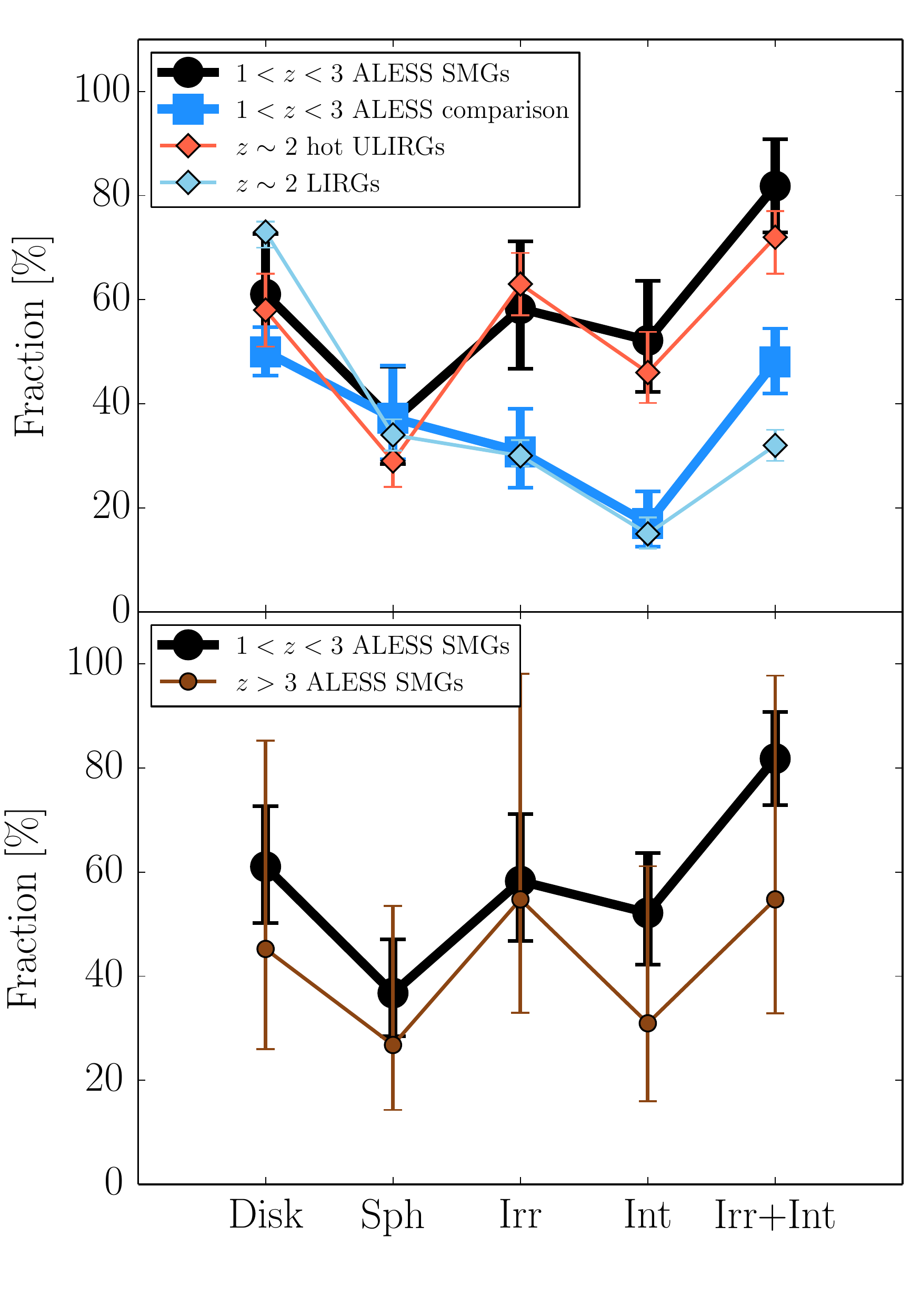}
       \caption{{\it Upper}: Fraction of the $H_{160}$-band detected ALESS SMGs at $z=$\,1--3, $z\sim$\,2 hot ULIRGs selected at 100/160\,$\mu$m, and $z\sim$\,2 LIRGs in each of the morphology classes (Disk, Disk; Sph, Spheroid; Irr, Irregular; Int, Interaction). The results for the $z\sim$\,2 hot ULIRGs and LIRGs are taken from \citealt{Kartaltepe:2012aa}. The Irr+Int class represents a combined class of irregular and interaction. Note that because the morphological classes are not mutually exclusive, the percentages do not sum to 100\%. {\it Lower}: Same as the upper panel but we now compare  SMGs at $z < $\,3 and those at $z > $\,3. The fraction of $z > $\,3 SMGs with disturbed morphologies is slightly lower than (but statistically comparable to) that of $z < $\,3 SMGs, confirming that our results of a high irregular fraction for SMGs are robust and not obviously affected by morphological $K$-correction.}
    \label{morphfract}
 \end{center}
\end{figure}

We compare our classifications to those from \citet{Kartaltepe:2012aa}, who presented similar visual classifications of a galaxy sample lying within the central region of the GOODS-S field and selected at 100 and 160\,$\mu$m using the PACS instrument \citep{Poglitsch:2010aa} onboard the {\it Herschel Space Observatory} ({\it Herschel}), as part of the GOODS-{\it Herschel} program \citep{Elbaz:2011aa}.  The ULIRGs presented by \citet{Kartaltepe:2012aa} have a mean $L_{\rm IR}$ of 2\,$\times$\,10$^{12} $\,L$_\odot$ and a mean redshift of $z=$\,2.2, {compared to the $z \sim $\,2 ALESS SMGs, $L_{\rm IR} \sim $\,3.0\,$ \times$\,10$^{12}$\,L$_\odot$ and $z=$\,2.1, the  ULIRG sample from \citet{Kartaltepe:2012aa} has a slightly lower mean $L_{\rm IR}$ but a similar mean redshift.}  These ULIRGs are also expected to be hotter in terms of dust temperature than the ALESS SMGs, and potentially more AGN-dominated, since they are selected at rest-frame $\sim$\,30--50\,$\mu$m. In addition \citet{Kartaltepe:2012aa}  present a non-{\it Herschel} selected comparison LIRG sample with redshift and $H_{160}$-band magnitude distributions matched to their ULIRG sample but lower infrared luminosity ($L_{\rm IR} <$\,3.2\,$ \times $\,10$^{11} $\,L$_\odot$), which is also plotted in Figure \ref{morphfract}.

We show this comparison in Figure~\ref{morphfract}, where we see that for all morphological classes, the  distributions of $z \sim $\,2 hot ULIRGs from \citealt{Kartaltepe:2012aa} are consistent with the $z \sim $\,2 ALESS SMGs, while the results for the ALESS SMGs differ significantly in the categories
related to disturbed morphology from those of LIRGs in \citet{Kartaltepe:2012aa}, as well as compared to our comparison low-luminosity field sample.  Hence, although at $z =$\,1--3 submillimeter-selected SMGs preferentially pinpoint the galaxies with lower dust temperature than the hot far-infrared-selected ULIRGs presented by \citet{Kartaltepe:2012aa}, it appears that the morphological mix seen in the ALESS SMGs is similar to that seen in comparably luminous PACS-selected ULIRGs at similar redshifts.

We next investigate the sensitivity of our classifications to the effects of redshift, in particular the  morphological $K$-correction: the variation in source morphology arising from color variations within galaxies, due to dust or age effects, coupled with the sampling of different restframe  wavelengths at different redshifts (e.g., \citealt{Kartaltepe:2014aa}). While we focus our analysis on $z < $\,3 SMGs, where the $H_{160}$-band fluxes more closely trace the bulk of the stellar emission, we can test whether the results of our visual classifications vary with redshift, in particular in the $z > $\,3 regime where the $H_{160}$-band imaging samples the rest-frame UV emission. It is possible that the rest-frame UV morphologies of $z > $\,3 SMGs will appear more irregular than that seen in the rest-frame optical/near-infrared sampling of the $z < $\,3 SMGs since the rest-frame UV is more sensitive to clumpy star-forming regions and dust obscuration.

%
%
\begin{figure*}
\begin{center}
    \leavevmode
      \includegraphics[scale=0.65]{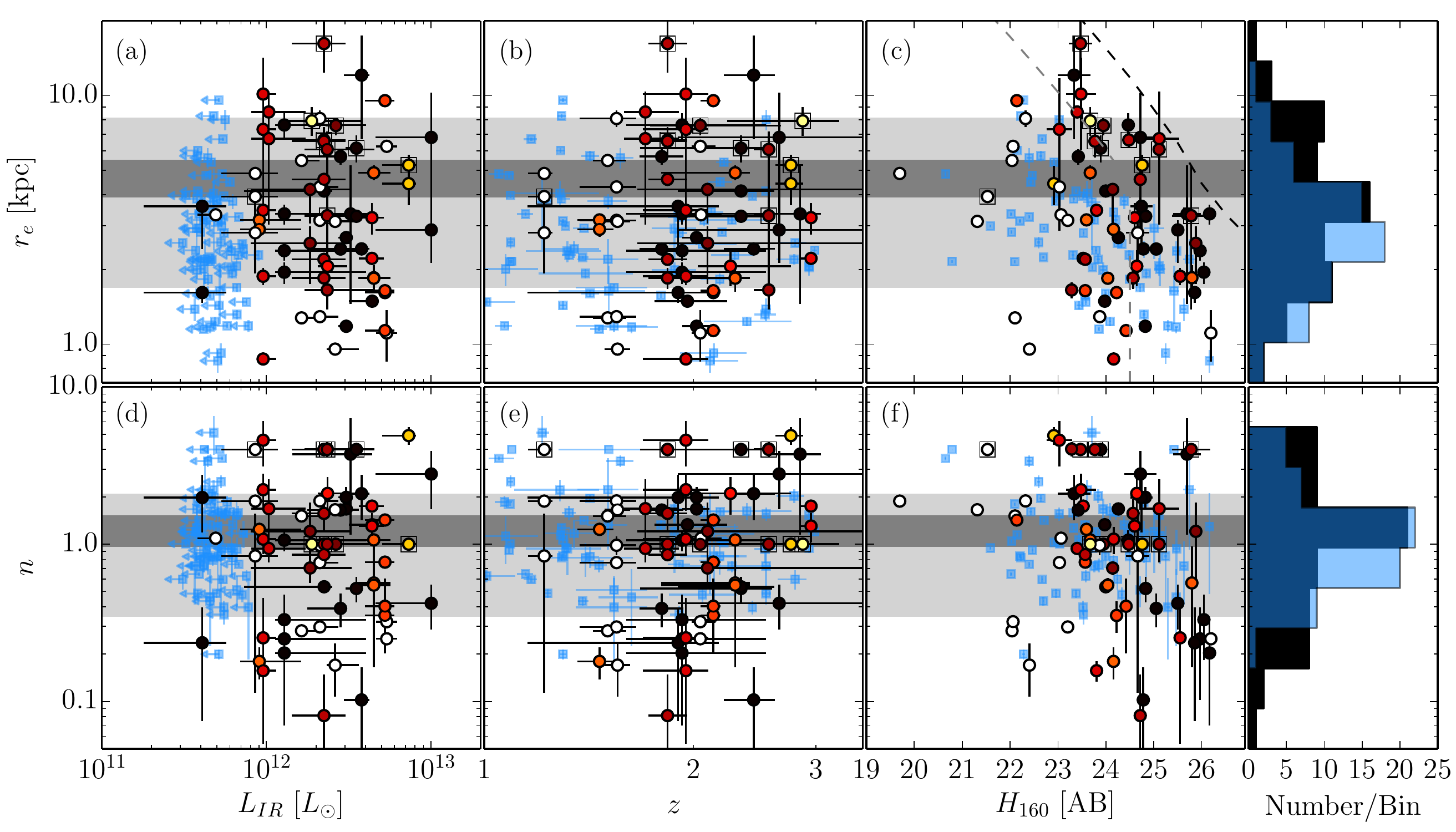}
       \caption{Panels (a)-(f) show the distribution of  effective radius ($r_e$) and S\'{e}rsic index ($n$) as a function of total luminosity ($L_{\rm IR}$; derived by \citealt{Swinbank:2014ul}), redshift ($z$; \citealt{Simpson:2014aa}, and the $H_{160}$-band AB magnitude for $z=$\,1--3 ALESS SMGs. The last panel of each row shows the histogram of either $n$ or $r_e$, in which black and blue represents ALESS SMGs and the comparison sample, respectively. The ALESS SMGs are color coded based on their $\chi^2$ values from the {\sc galfit} model fit (see \autoref{tab:galfit}), and those enclosed by an extra black square are those with a fixed S\'{e}rsic index. The results of the comparison sample are shown in blue. The grey (black) dashed curves in panel (c) represent the maximum $r_e$ detectable based on \autoref{eqa:sersicfin} at the sensitivity of either the previous NICMOS studies or our new  $H_{160}$-band imaging, respectively. The dark grey horizontal bands are the median values of $H_{160}$-band components with $H_{160} \leq $\,24 for ALESS SMGs (unbiased $r_e$ at $<$\,15\,kpc; also see text in \autoref{subsec:vis}) with the width showing the bootstrapped errors. The light grey horizontal bands represent the 1\,$\sigma$ intrinsic scatter. There is no detectable variation in $n$ or $r_e$ with $L_\text{IR}$, $z$, and $H_{160}$ for the ALESS SMGs from our the $H_{160}$-band imaging. While the sources in the comparison sample are consistent with the ALESS SMGs in terms of $n$ (1.0\,$\pm$\,0.1 versus 1.2\,$\pm$\,0.3 respectively),  on average they have a smaller $r_e$ than the SMGs (2.5\,$\pm$\,0.3\,kpc versus 4.4$^{+1.1}_{-0.5}$\,kpc).}
    \label{fig:ren}
 \end{center}
\end{figure*}

In the bottom panel of Figure \ref{morphfract} we investigate the fraction of SMGs in each morphological class in high- and low-redshift subsamples. Although suffering from small number statistics in the $z > $\,3 subsample, the fraction of $z > $\,3 SMGs with disturbed morphologies is slightly lower than (but statistically consistent with) that of lower-redshift SMGs. This is mostly due to the $H_{160}$--$z$ relation in Figure \ref{hmag}, which shows that higher redshift SMGs have lower $H_{160}$-band fluxes. This trend results in many $z > $\,3 SMGs appearing as a single, faint source in the $H_{160}$-band imaging, and they are unlikely to be classified as a clumpy irregular. The lower fraction of irregulars at $z > $\,3 is thus likely to be due to the depth of the available imaging, and deeper images are needed to investigate the true morphology of $z > $\,3 SMGs. On the other hand, our results of a high irregular fraction at $z < $\,3 are robust and not affected by morphological $K$-correction.

It is also worth noting that the presence of AGN in the sample does not seem to affect our conclusions.
There are six ALESS SMGs that are identified as AGNs through X-ray detections (these are noted  in \autoref{tab:galfit}; \citealt{Wang:2013aa}), with five of these  at $z=$\,1--3. However, we do not observe a significant difference in the visual classes between AGN SMGs and non-AGN SMGs. 

\subsection{S\'{e}rsic index and Effective Radius}\label{subsec:vis}
In \autoref{fig:ren} we plot the S\'{e}rsic index ($n$) and the effective radius ($r_e$) for every component of the $z = $\,1--3 ALESS SMGs detected in our $H_{160}$-band imaging as a function of infrared luminosity, redshift, and $H_{160}$-band magnitude. 

Overall there are no obvious trends in any combination of parameters for the ALESS SMGs, regardless of the goodness of the fit ($\chi^2_\nu$). However, systems with higher $\chi^2_\nu$ tend to be at lower redshifts and have brighter $H_{160}$-band fluxes, suggesting that low surface brightness features such as tidal tails or clumpy structures, which are only revealed in these closer and brighter systems, drive the higher values of $\chi^2_\nu$. Indeed, by adding noises into the thumbnails for these high $\chi^2_\nu$ systems and rerunning the {\sc galfit} analysis we found consistent best fit solutions but with {lower} $\chi^2_\nu$ values. We therefore concluded that these bright, low redshift but high-$\chi^2_\nu$ systems are not intrinsically different in $n$ and $r_e$ from those at higher redshift with fainter $H_{160}$-band fluxes.

On the other hand, an absence of sources with large $r_e$ and faint-$H_{160}$ components can be seen in \autoref{fig:ren}, in the following we argue that this is due to the sensitivity of the $H_{160}$-band imaging.

We can re-write the S\'{e}rsic profile (\autoref{eqa:sersic}) in a simpler form as

\begin{equation}
\Sigma(r) = \Sigma_o\exp\left[-\kappa\left(\frac{r}{r_e}\right)^{1/n}\right]
\label{eqa:sersicsimp}
\end{equation}

\noindent
where $\Sigma_o = \Sigma_ee^{\kappa}$ is the central surface brightness, and the total flux would be

\begin{equation}
F=\int_0^{2\pi}d\phi\int_0^\infty r\Sigma(r)dr = 2\pi\Sigma_o \int_0^\infty r\exp\left[-\left(\frac{r}{\alpha}\right)^{1/n}\right]dr
\label{eqa:sersicint}
\end{equation}

\noindent
where $\alpha = r_e/\kappa^n$. We then make a substitution of $t = (r/\alpha)^{(1/n)}$, and thus $r=t^n\alpha$ and $dr = \alpha nt^{n-1}dt$. Finally \autoref{eqa:sersicint} becomes

\begin{equation}
F= 2\pi\Sigma_o\alpha^2 n\int_0^\infty t^{2n-1}\exp\left(-t\right)dt
\label{eqa:sersicintsht}
\end{equation}

\noindent
and the integral is actually a Gamma function ($\Gamma(y) = \int_0^\infty x^{y-1}\exp\left(-x\right)dx$). Thus \autoref{eqa:sersicintsht} becomes

\begin{equation}
F= 2\pi\Sigma_o\alpha^2 n\Gamma(2n)
\label{eqa:sersicsub}
\end{equation}

\noindent
and 

\begin{equation}
\Sigma_o=\frac{F}{2\pi\alpha^2 n\Gamma(2n)}
\label{eqa:sersicsig}
\end{equation}

\noindent
We then substitute this back into \autoref{eqa:sersicsimp} and we now have

\begin{equation}
r_e^2 \exp\left[\kappa\left(\frac{r}{r_e}\right)^{1/n}\right] = \frac{\kappa^{2n}F}{2\pi n\Gamma(2n)\Sigma(r)}
\label{eqa:sersicfin}
\end{equation}

\noindent
\autoref{eqa:sersicfin} means that, taking our deepest surface brightness sensitivity ($\Sigma(r_{det})$) 
and the total number of pixels required for the {\sc SExtractor} detection threshold, we can determine the radius ($r _{det}$) corresponding to any total flux $F$ and so solve for the maximum detectable $r_e$ given any $n$ ($\kappa$ is a function of $n$). In practice, for a given $F$ ($H_{160}$ in our case), we solved for the maximum $r_e$ using \autoref{eqa:sersicfin} for a range of S\'{e}rsic profile ($n = $\,4, 2, 1, 0.5, 0.1, and the corresponding $\kappa = $\,7.67, 3.67, 1.68, 0.70, 0.02, respectively), and converted $r_e$ to kpc by adopting the mean redshift $z = $\,2.1. We plot the results in \autoref{fig:ren}. This selection line bounds the parameter space of our observations  and demonstrates that for a given detection limit there is a size bias toward smaller $r_e$ when detecting  components  close to the detection limit. Thus one needs to be cautious when comparing typical sizes between two samples if they were observed to different depths. 

The median effective radius (or half-light radius; $r_e$) of all $H_{160}$-band components in \autoref{tab:galfit} is 3.3\,$\pm$\,0.2 kpc (all the errors  quoted on median values in this paper were obtained through a bootstrapping method). However, if we only take components that have $H_{160} \leq $\,24, which should be complete for measured sizes up to $\sim$\,15\,kpc, we obtained a median $r_e$ of 4.4$^{+1.1}_{-0.5}$\,kpc (dark grey horizontal band in the top panels of \autoref{fig:ren}) with an intrinsic scatter of the $H_{160}$-band component sizes of 1.7--8.1\,kpc. Applying a $H_{160}$-band cut at $H_{160} \leq $\,24 provides a fairer estimate of the rest-frame optical sizes for ALESS SMGs, assuming that the $r_e$ distribution at $H_{160} > $\,24\,mag is not drastically different than that at $H_{160} \leq $\,24\,mag.   Note that our measurements of the  median $r_e$ are not sensitive to the $H_{160}$ cut between 22 and 24\,mag.  
 
A NICMOS study by \citet{Swinbank:2010aa} of the $H_{160}$-band morphologies of a sample of 25 radio-identified SMGs determined a median $r_e$ of 2.8\,$\pm$\,0.4\,kpc, consistent with our results based on all $H_{160}$-band components but somewhat lower than our unbiased measurements with a $H_{160}$ cut \footnote{We should note that \citet{Swinbank:2010aa} defines the half-light radius ($r_h$) as the radius at which the flux is one-half of that within the Petrosian radius. Based on \citet{Graham:2005aa} a profile with $n \le $\,1 $r_h$ equals to $r_e$, while $r_h$ is only $\sim$\,70\% of $r_e$ for $n = $\,4. As most SMGs have low $n$, we do not expect $r_h$ to be significantly different than $r_e$. Indeed, we measured $r_h$ on our SMGs with $H_{160} \leq $\,24\,mag following the procedure of \citet{Swinbank:2010aa}, and find a median $r_h$ of 4.3\,kpc.}. However, considering the shallower sensitivity of NICMOS imaging ($\mu_H \sim $\,24.5\,mag arcsec$^{-2}$, $1\,\sigma$), we might expect a bias toward low $r_e$. Indeed, we show the NICMOS $r_e$ sensitivity as the dashed grey curve in panel (c) of \autoref{fig:ren}. The median $r_e$ of our $H_{160}$-band components that meet the NICMOS sensitivity is 3.2\,$\pm$\,0.5\,kpc, consistent with the results of \citet{Swinbank:2010aa}, which we conclude were biased down somewhat by the surface brightness limit of their study.

Recently, \citet{Targett:2013aa} used the CANDELS $H_{160}$-band imaging to study the morphology of a sample of a mixture of millimeter and submillimeter sources uncovered by AzTEC at 1.1\,mm and LABOCA at 870\,$\mu$m (the latter taken from the same LESS survey used as the basis of ALESS). They found that their sample of 24 candidate counterparts to the submillimeter sources  has a mean $r_e$ of 4.3\,$\pm$\,0.5\,kpc, consistent with our results. With a comparable $H_{160}$-band  depth   to our study, it is perhaps not surprising that both groups found a consistent median $r_e$. However, without high-resolution follow-up observations with interferometers, their results are very likely to suffer from the contamination by mis-identifications ($\sim$\,20\%; \citealt{Hodge:2013lr}), for example the ALMA observations show that their proposed counterpart to LESS\,J033217 is not the SMG. In addition, {the selection function for this sample is complicated} by the use of  a heterogeneous submillimeter and millimeter sample, where the different depths, resolutions and selection  wavelengths will bias the sample in different ways in redshift and/or dust temperature.  

The surface brightness sensitivity of our WFC3 imaging tends to bias the mean $r_e$ of the ALESS SMGs toward artificially smaller values at faint magnitudes. However, as shown in panel (c) of \autoref{fig:ren}, the $H_{160}$-band imaging is actually sufficiently deep to reveal a trend between $r_e$ and $H_{160}$ for the field sample, in which $d(log(r_e))/d(H_{160}) \sim -0.2$. Given {that} there is no correlation between $n$ and $H_{160}$ for the comparison sample (panel (f) in \autoref{fig:ren}), based on \autoref{eqa:sersicsub} the trend is consistent with the scenario that the central surface brightness ($\Sigma_0$) of the comparison field sample remains constant on average across the $H_{160}$ range and the apparent size of the source then determines its integrated brightness. Perhaps more interestingly, we can use the deeper WFC3 data to test the influence of the NICMOS sensitivity limits on the comparison of the field and SMG samples from \citet{Swinbank:2010aa}.  
We find that while the median $r_e$ of the whole comparison field sample is 2.5\,$\pm$\,0.2\,kpc, the median $r_e$ is 3.1\,$\pm$\,0.3\,kpc for those sources that meet the NICMOS sensitivity criteria in \citet{Swinbank:2010aa}.   Hence if we had to rely on shallower NICMOS imaging, we would have concluded that the ALESS SMGs and the low-luminosity field comparison sample are  indistinguishable in $r_e$ (this is consistent with the findings in \citet{Swinbank:2010aa}). However, employing the full depth of our WFC3 imaging we can begin to separate these  two populations in terms of their typical sizes and see that the SMGs are larger on average than the general field population at these redshifts. 

Turning now to the distribution of S\'{e}rsic index ($n$) in \autoref{fig:ren}, we can see that these are highly peaked at low $n$ for both the ALESS SMGs and the comparison sample, with $\sim$\,80\% of the sources having $n <$ 2. Indeed, for the ALESS SMGs the median $n$ for those $H_{160}$-band components with $H_{160} \leq$ 24\,mag is 1.2\,$\pm$\,0.3, with a 1\,$\sigma$ dispersion of 0.4--2.1. The median $n$ is not sensitive to the $H_{160}$ cut (the median $n$ is 1.0\,$\pm$\,0.2 for all the $H_{160}$-band components) but we make the cut at $H_{160} \leq 24$\,mag just to be consistent with the measurement of the median $r_e$. The median $n$ for the sources in the comparison field sample is 1.0\,$\pm$\,0.1. Since there is no apparent bias in the derived S\'{e}rsic index as a function of image depth (indeed we find solutions for a wide range of $n$ in \autoref{eqa:sersicfin} at $H_{160} < $\,27\,mag), it is not surprising that the median $n$ from \citet{Swinbank:2010aa} and \citet{Targett:2013aa}, 1.4\,$\pm$\,0.8 and 1.0, respectively, are consistent with our results. 

In this analysis we have assumed that every $H_{160}$-band component which overlaps the ALMA synthesized beam is a counterpart of the ALESS SMG. While in most cases this is likely to be true, we do not have redshift measurements to confirm our assumptions, in particular for a few ambiguous cases (e.g., ALESS\,10.1 or ALESS\,92.2), potentially casting doubts on the robustness of our identifications. We test whether a few ambiguous cases would affect our results by just measuring the most likely component of each ALESS SMG (labelled H1 in \autoref{tab:galfit}) \footnote{In all cases but ALESS\,114.2, these are the closest components to the ALMA positions. {The $H_{160}$-band counterpart of ALESS\,114.2 can be modelled with two components (as can be seen in \autoref{tab:galfit} as well as in \autoref{postage} in the Appendix): H1 represents the main component accounting for the bulk of the $H_{160}$-band flux; H2 models the lower level emission surrounding H1. Although H1 has a slightly larger offset to the ALMA position, H1 is likely to be the representative $H_{160}$-band component for ALESS\,114.2.}}. We found the median $r_e$ and $n$ unchanged, with $r_e =$\,4.1$^{+1.6}_{-0.8}$\,kpc and $n =$\,1.2$^{+0.2}_{-0.3}$, respectively. However, we found the intrinsic scatter for $r_e$ to be slightly smaller, 2.7--7.7\,kpc, which is likely to be the lower limit to the true scatter as our test did not take into account some secondary components which are likely to be the counterparts of the ALESS SMGs (i.e., ALESS\,49.2, ALESS\,57.1, ALESS\,79.2, ALESS\,87.1). Similarly, the scatter based on all the $H_{160}$-band components as presented earlier is likely to be the upper limit as a few may not be the correct counterparts. We adopted the upper limit of the $r_e$ scatter to maintain a consistent approach throughout this paper. 

\subsection{Positional Offsets}\label{sec:offsets}

%
%
\begin{figure}
\begin{center}
    \leavevmode
      \includegraphics[scale=0.68]{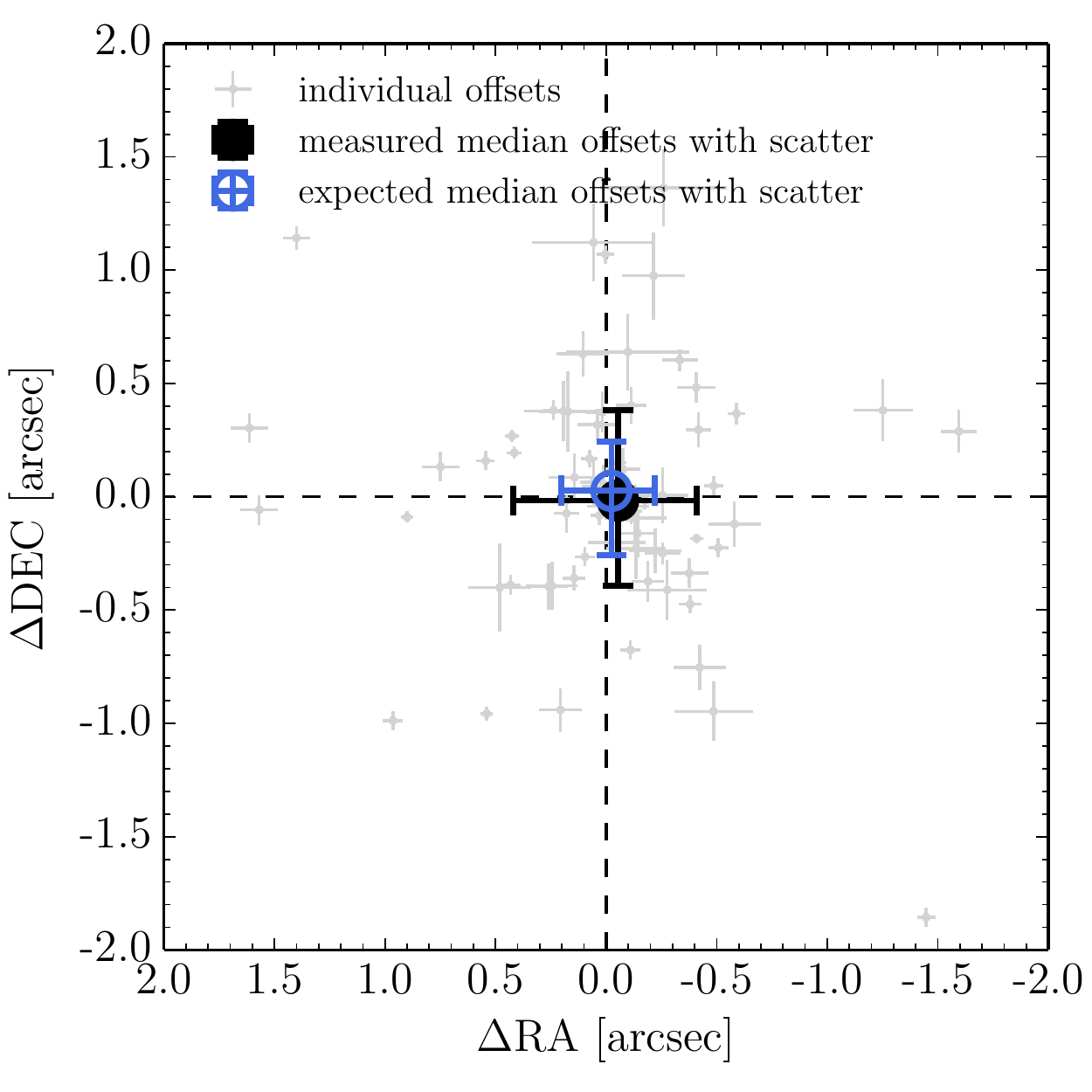}
       \caption{The offsets in R.A.\ and Dec.\ between the $H_{160}$-band component and the corresponding ALMA 870\,$\mu$m emissions ($H_{160}
-$\,ALMA; as quoted in \autoref{tab:galfit}). The individual component is shown in grey with errors estimated based on the uncertainties from both {\sc galfit} and the ALMA observations. The measured (expected) median offsets are shown in black (blue) circle with the symbol size corresponding to about two times the error ($-0\farcs05\pm0\farcs05$ in R.A.\ and $0\farcs02\pm0\farcs06$ in Dec.). The intrinsic scatters are shown in black error bars, which on average are $0\farcs4\pm0\farcs05$. The significantly larger measured positional scatter compared to that expected suggests that most of the dusty star-forming regions are not located close to the center of the stellar distribution as seen in the restframe optical.}
    \label{fig:offsets}
 \end{center}
\end{figure}
With each $H_{160}$-band component of our ALESS SMGs reliably modelled using {\sc galfit} with good positional constraints, as well as the sub-arcsecond positional accuracy provided by the ALMA imaging, we can investigate the positional relation between the dusty star-forming regions traced by ALMA and the stellar components traced by our $H_{160}$-band imaging. 

In \autoref{fig:offsets} we plot the positional offsets in R.A.\ and Dec.\ between all our $H_{160}$-band components at $z = $\,1--3 in \autoref{tab:galfit} and their corresponding ALMA 870\,$\mu$m positions (defined as $H_{160}-$\,ALMA). In addition to the {\sc galfit} errors, for each $H_{160}$-band component we folded in the positional errors of the original ALMA observations derived by \citet{Hodge:2013lr}. We measured a median $\Delta$\,R.A.\ of $-0\farcs05\pm0\farcs05$ and a median $\Delta$\,Dec.\ of $-0\farcs02\pm0\farcs06$. We also measured an average intrinsic scatter of $0\farcs40\pm0\farcs05$. 

As we included in this analysis any $H_{160}$-band component that intersects the ALMA synthesized beam ($1\farcs6$ FWHM) as potentially part of the SMG, it is possible that some fraction of the components are not associated. Hence if we restricted our analysis to just those components whose centroids fall within the ALMA beam we obtained an average intrinsic scatter of $0\farcs34\pm0\farcs03$, and median offsets of $-0\farcs07\pm0\farcs05$ in R.A.\ and $-0\farcs03\pm0\farcs06$ in Dec.

The systematic offsets from both definitions of the counterparts are not significant, confirming our initial alignment of the astrometric frames described in \autoref{sec:hst}.  As stated in \autoref{sec:ooda} the intrinsic scatter between the IRAC 3.6\,$\mu$m sources and their calibrated $H_{160}$-band counterparts is 0$\farcs$15--0$\farcs$17. We expect the scatter between ALMA and IRAC to be comparable to this, {meaning that} if the stellar components are well-aligned with the dusty star-forming regions, we derive that the expected scatter between ALMA and IRAC is $\sqrt2\times 0\farcs17 = 0\farcs24 $, adopting the higher scatter to be conservative. The measured scatter for the ALESS SMGs is $0\farcs4\pm0\farcs05$, significantly larger than the expected value, $0\farcs24$. This is still the case if we only consider the scatter obtained from the $H_{160}$-band components located within the typical ALMA beam. In fact, only $\sim$\,34\% (44\%) of the data points (those within the synthesized ALMA beam) are consistent with no offset. 

A significantly larger scatter in positional offsets between the $H_{160}$-band components and the corresponding ALMA 870\,$\mu$m peaks suggests that majority of the dusty star-forming regions are not located close to the center of the stellar distribution seen in the restframe optical.  This could simply suggest that the dusty star-forming regions are so obscured that the rest-frame optical imaging does not reflect the true stellar morphology. We tested this scenario by comparing the measurements between lower ($z=$\,1--2) and higher ($z=$\,2--3) redshift subsamples. If obscuration is truly a significant factor and the dusty star-forming regions are indeed located close to the center of the stellar distribution, we would expect the scatter for the low-redshift subsample to be smaller, as the $H_{160}$-band imaging probes their rest-frame 5000--8000$\AA$ and is less obscured by dust. However, we found no statistical difference in the positional scatter between the low-redshift and high-redshift subsample. We therefore conclude that it is more likely that the offset between the $H_{160}$-band components and the corresponding ALMA 870\,$\mu$m peaks reflects real misalignment between the dusty star-forming regions and the locations of the majority of the stellar masses within the SMGs.

\section{Discussion}\label{sec:dis}

\subsection{Massive Disks or Mergers?}
Recently, a few studies have analysed the near-infrared morphologies of SMGs in order to shed light on the stellar distribution of these high-redshift, dusty star-forming sources \citep{Swinbank:2010aa, Targett:2013aa, Wiklind:2014aa}. Various quantitative analysis have been conducted, including CAS method, the Gini/M20 parameters, as well as the S\'{e}rsic indices, and all the studies have claimed that SMGs are  massive disks \citep{Targett:2013aa} and no more likely to appear as major mergers in the rest-frame optical than those more typical star-forming galaxies selected in rest-frame UV \citep{Swinbank:2010aa}. It has also been claimed that SMGs that are selected at 870\,$\mu$m represent a more isolated, heterogeneous population \citep{Wiklind:2014aa} in contrast to 100/160\,$\mu$m selected sources at similar redshifts \citep{Kartaltepe:2012aa}.

In this study we have discovered that although our $H_{160}$-band detected ALESS SMGs appear disk-like in quantitative analysis (median $n$ of 1.2\,$\pm$\,0.3), most ($\sim$\,80\%) are in fact visually classified as either irregulars or interacting systems. A lack of visual classification on a statistical sample seems to be the one reason that  previous studies mistakenly concluded that most SMGs are disks. After all, the non-parametric tools such as CAS and Gini/M20 have limited power to differentiate irregulars/interacting systems from disks \citep{Swinbank:2010aa, Wiklind:2014aa}. 

Now, the question becomes whether sources appearing as irregulars or interacting systems are clumpy rotating disks or mergers. Recently, many efforts have been made to shed light on the triggering mechanism of star formation through studies of gas dynamics. Kinematic studies using  integral field unit (IFU) spectroscopic observations of ionised atomic lines (H$\alpha$ or [O{\sc iii}]) from $z \sim $\,1--3 star-forming galaxies have found that some sources that appear as irregulars in the optical/near-infrared are rotationally dominated (e.g., \citealt{Genzel:2011aa}), {although signs of rotationally-supported dynamics do not preclude mergers as the merger signatures could be lost due to limited spatial resolution of high-redshift sources (e.g., \citealt{Goncalves:2010aa})}. However, most of these studies were focused on rest-frame UV selected star-forming galaxies that have lower SFRs than SMGs. Recent, IFU studies of small samples of SMGs, including one of our sources ALESS\,67.1, have instead found either a lack of evidence for global ordered rotation, or that most SMGs are classified as mergers based on a kinemetric analysis of the velocity and dispersion field asymmetry \citep{Alaghband-Zadeh:2012aa, Menendez-Delmestre:2013aa}. These results are consistent with an earlier IFU study on a sample of six SMGs by \citet{Swinbank:2006aa}, which shows that these sources have distinct dynamical subcomponents, suggesting a merging/interacting nature of these systems. 
Equally, a dynamical study on ALESS\,73.1, using high-resolution (0$\farcs$5) interferometric observations with ALMA of the [C{\sc ii}]\,157.7$\mu$m fine structure  line, has found that the gas reservoir traced by this emission has kinematics which are consistent with a rotating disk \citep{De-Breuck:2014aa} and similar gas disks have been claimed  in some other high-redshift ULIRGs (e.g., \citealt{Swinbank:2011aa, Hodge:2012fk}). 
Again this doesn't preclude a merger origin as a gas disk is expected to reform during  the final coalescence of a gas-rich merger  (e.g., \citealt{Robertson:2006aa, Hopkins:2009aa}). 

In any case, the mixed results from studies of gas kinematics highlight the complications involved in explaining the dynamical history of SMGs based solely on the observations of molecular or ionised gas. In the following we study the triggering mechanisms of SMGs through the evolution of S\'{e}rsic index and effective radius based on our results on the collisionless stellar components of the ALESS SMGs.

\subsubsection{Evolution of Size and S\'{e}rsic index}\label{sec:size}
Due to their high star-formation rates (SFRs), bright SMGs are expected to be short lived, with a lifetime of $\sim$\,100\,Myr (e.g., \citealt{Chapman:2005p5778}), which is similar to the estimated lifetimes based on SMG clustering (e.g., \citealt{Hickox:2012kk}). Indeed, the median SFR of our ALESS SMGs is $\sim$\,500\,$M_\odot$\,yr$^{-1}$ (3\,$ \times $\,10$^{12}$\,L$_\odot$), and for a typical SMG gas mass of $\sim $\,5\,$ \times$\,10$^{10} $\,M$_\odot$ (e.g., \citealt{Bothwell:2013lp}), the time for ALESS SMGs to exhaust their gas reservoir is of order 100\,Myr. If the quenching of star formation indeed happens shortly after the SMG phase, then the likely descendants of our $z \sim $\,2 ALESS SMGs would be $z \sim $\,2 quiescent galaxies. Such connections between $z = $\,2--3 SMGs and $z \sim $\,2 quiescent galaxies have been made by other studies using similar arguments on SFRs and stellar mass growth (e.g., \citealt{Wardlow:2011qy, Fu:2013mm, Toft:2014aa}). Unfortunately, while this is a conceptually tidy proposal, 
there are potential problems with the details of how the transition is made,  in particular from the point of view of transforming the stellar distributions. With a well-defined and reliably identified sample of SMGs at $z \sim $\,2, we can test this proposal and investigate  different quenching mechanisms.

\begin{figure}
\begin{center}
    \leavevmode
      \includegraphics[scale=0.48]{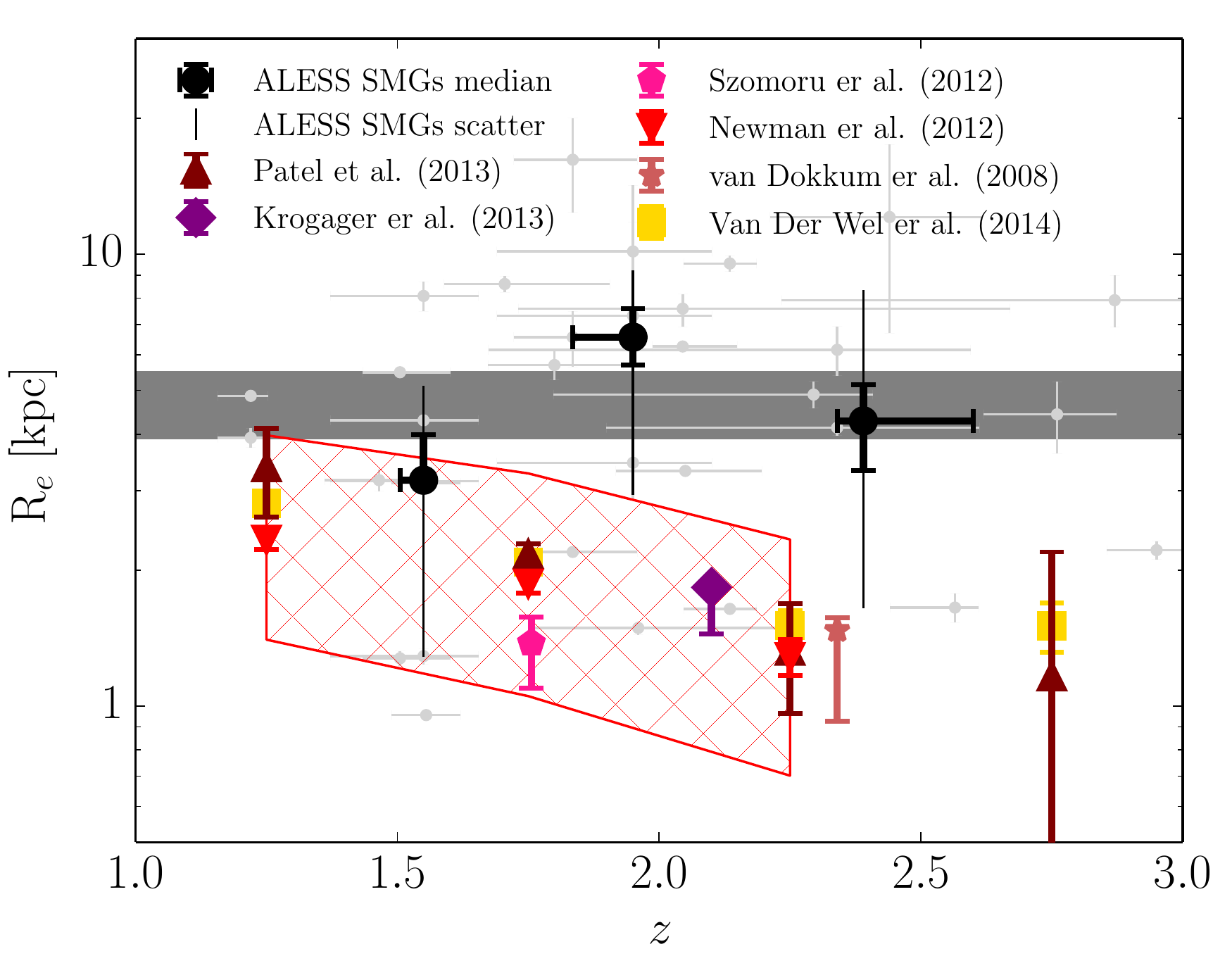}
       \caption{A plot of redshift versus effective radius ($r_e$) for different galaxy populations. The results from the individual $H_{160}$-band component of the ALESS SMGs with $H_{160} <$\,24 are plotted in light grey, with the three equal-sized bins shown in black points and the median size of 4.4\,$\pm$\,0.2\,kpc in dark grey. The intrinsic scatter in each bin is plotted as vertical black lines. Various measurements from the literature for the sizes of  $z\sim$\,2 quiescent galaxies are shown  with the typical scatter illustrated by the red grids from  \citealt{Newman:2012aa}. The SMGs are typically 2--3\,$\times$ larger than $z\sim$\,2 quiescent galaxies at a similar epoch.}
    \label{rez}
 \end{center}
\end{figure}

In \autoref{rez} we show the distribution of effective radius ($r_e$) versus redshift for the $H_{160}$-band components of the ALESS SMGs at $z = $\,1--3 with $H_{160} \leq $\,24\,mag. We also show the median in three equal number bins (11 SMGs in each) with redshift. As we discussed in \autoref{subsec:vis}, there is no apparent size evolution for the ALESS SMGs. The median size for the ALESS SMGs at $z = $\,1--3 ($r_e = $\,4.4$^{+1.1}_{-0.5}$\,kpc) is shown as the dark horizontal band in \autoref{rez}, and we also plot the size measurements of $z \sim $\,2 quiescent galaxies made using the NIC2 camera \citep{van-Dokkum:2008aa}, and with the WFC3 in CANDELS imaging in COSMOS \citep{Krogager:2013aa}, GOODS-S \citep{Szomoru:2012aa}, and UDS field \citep{Newman:2012aa, Patel:2013aa}. {We also include the latest size measurements on apparently passive galaxies at $z < $\,3 based on  the combined CANDELS and 3D-{\it HST} surveys \citep{van-der-Wel:2014aa}. All sizes were converted to the $r_e$ measured along the major axis. The conclusions remain unchanged if all $r_e$ were instead converted to the commonly-used circularized effective radius, $r_{e,circ} = r_e\sqrt{b/a}$, where $b/a$ is the projected axis ratio.}

%
%

The size difference between SMGs and quiescent galaxies is immediately apparent from \autoref{rez}, especially at $z > $\,2 where ALESS SMGs are significantly larger than the quiescent galaxies. A large range in size  is generally reported for $z =$\,2--3 quiescent galaxies (e.g., \citealt{Szomoru:2012aa, Newman:2012aa, van-der-Wel:2014aa}), with a typical scatter of about 0.2--0.25 dex for mass-selected samples with $\ge $\,5\,$\times$\,10$^{10} M_\odot$ (the completeness limit for CANDELS imaging at $z < $\,3). A few $z = $\,2--3 quiescent galaxies are found to be located within the size range of SMGs, making them candidate descendants of SMGs. However, considering the comoving density of $M_\star > 5\times10^{10} M_\odot$ quiescent galaxies at $z \sim $\,2, $n_c \sim 10^{-4}$\,Mpc$^{-3}$ \citep{Newman:2012aa, Patel:2013aa}, and the fraction of them with $r_e > $\,1.7\,kpc (the lower limit of the ALESS SMGs), $\sim$\,10--40\% \citep{Trujillo:2007aa, Newman:2012aa, Szomoru:2012aa, van-der-Wel:2014aa}, the comoving density of $z \sim 2$ quiescent galaxies with $r_e > $\,1.7\,kpc is only $\sim $\,1--4\,$\times$\,10$^{-5}$\,Mpc$^{-3}$. Given the duty cycle, $\sim$\,10  Gyr$^{-1}$, and the comoving number density, $\sim $\,10$^{-5}$\,Mpc$^{-3}$, of the ALESS SMGs at $z = $\,2--3 \citep{Simpson:2014aa}, these large $z \sim $\,2 quiescent galaxies would only comprise $\sim$\,10--40\% of the SMG descendants at $z \sim $\,2, if they are indeed on the same evolutionary track. This means that if they are to become typical $z\sim$\,2 quiescent galaxies the majority of $z > $\,2 SMGs have to go through a transformational phase that significantly reduces their stellar half-light radii, as well as increasing their S\'{e}rsic indices ($n \sim $\,1 to $\sim$\,4), before being quenched. Note that the smaller scatter in the sizes, $\sigma_{re} = $\,1.7\,kpc, remains unchanged even if we remove the $H_{160}$ magnitude limit on the ALESS sample used for the size measurements. It is also worth emphasising that this lower bound on the size distribution of SMGs  is a conservative limit, as presented in \autoref{subsec:vis}, and the true lower bound on the sizes is likely to be higher, suggesting that more than 60\% of $z=$\,2--3 SMGs need to go through a significant transitional phase of the stellar distribution.  

For any mechanism that drives the transformation of the stellar distribution, it must  increase the central stellar masses by either rearranging the stellar mass distribution or by forming a significant amount of young stars through bursts of centrally concentrated star formation, or both. A significant intrinsic offset between the dusty star-forming regions and the stellar components of ALESS SMGs, as shown in \autoref{sec:offsets}, means that if ALESS SMGs are isolated, secularly evolving disks, the bulk of the newly formed young stars from the current epoch of star formation need to migrate toward the central regions, which is unlikely given the collisionless nature of the stellar components. On the other hand, recent theoretical work by \citet{Dekel:2014aa} suggests that at $z \sim $\,2 rich inflows of pristine gas from the intergalactic medium (IGM) trigger violent disk instability and dissipatively drive gas into the center, shrinking the gaseous disk into a compact stellar distribution through on-going star formation, and further evolving into compact quiescent galaxies with high S\'{e}rsic indices through various quenching mechanisms. {This idea is similar to the bulge formation through migration of long-lived giant clumps in gas-rich disks (e.g., \citealt{Elmegreen:2008aa, Perez:2013aa, Bournaud:2014aa})}. However, in order for the inflow of gas to shrink through dissipative processes, the star formation timescale needs to be longer than the inflow timescale, which \citet{Dekel:2014aa} predict to be $\sim$\,250\,Myr, to avoid turning the majority of the gas into stars before it reaches the center. Unfortunately, the prediction of a long star-formation timescale is inconsistent with recent spectroscopic studies of $z \sim $\,2 quiescent galaxies, in which spectral line modelling yield stellar populations that have undergone a fast quenching star-formation history (SFH) with an $e$-folding timescale of $\tau < $\,250\,Myr, in the typical exponentially declining SFHs \citep{Kriek:2009aa, van-de-Sande:2013aa, Krogager:2013aa}. Furthermore, for $z\sim$\,2 quiescent galaxies, rapidly quenching SFHs not only weaken the likelihood that their progenitors are gaseous disk galaxies and/or compact star-forming galaxies (\citealt{Barro:2013aa}), but also strengthen their connections to $z\sim$\,2 SMGs. A similar link between quiescent galaxies and SMGs has also been suggested at $z > $\,3 based on the short ($<$\,100\,Myr) star formation timescale \citep{Marsan:2014aa}. In summary, given the estimated short lifetime of SMGs ($\sim$\,100\,Myr), it appears unlikely that the majority of the ALESS SMGs at $z = $\,2--3 can be significantly transformed into quiescent galaxies with a de Vaucouleur stellar profile through disk instability. 

On the other hand, major galaxy mergers have been shown to efficiently transform stellar distributions from disk-like to de Vaucouleur's profile through tidal forces and transfer of angular momentum (e.g., Barnes 1988; Barnes \& Hernquist 1996). Recent hydrodynamical simulations have shown that this is also the case for high-redshift gas-rich mergers \citep{Hopkins:2013aa}. Based on \autoref{eqa:sersicsig}, by fixing the total $H_{160}$-band fluxes and adopting a typical central surface brightness ($\Sigma_o$) of $z \sim $\,2 quiescent galaxies (median $\Sigma_o \sim $\,17.8\,mag\,arcsec$^{-2}$; \citealt{Szomoru:2012aa}), a simple transformation of S\'{e}rsic indices from $n=$\,1 to $n=$\,4 would make the effective radius of a source decrease from 4.4\,kpc to $\sim$\,1\,kpc. If the current burst of star formation doubles the total $H_{160}$-band flux then the $r_e$ would be $\sim$\,1.4\,kpc, consistent with the size of $z \sim $\,2 quiescent galaxies. Indeed, together with recent IFU dynamical studies, our findings of a high fraction of systems with disturbed stellar morphologies ($\sim$\,80\%) and significant offsets between star-forming regions and the stellar components suggest specifically that the majority of $z=$\,2--3 SMGs are likely to be early/mid-stage mergers.

\section{Summary}\label{sec:sum}
We have analysed $H_{160}$-band imaging taken with the WFC3 camera mounted on the {\it HST}, of a sample of 48 ALMA-identified  SMGs from the ALESS survey, as well as a comparison sample of 58 ALMA-undetected normal field galaxies at $z=$\,1--3 located within the ALMA primary beam. The sample of SMGs is  drawn from an interferometric follow-up study of a flux-limited SMG sample uncovered by {the single-dish LABOCA survey at 870\,$\mu$m in ECDFS (LESS survey; \citealt{Weis:2009qy})}. The key results from our study are:

\begin{enumerate}
\item We found that 38 out of 48 ALESS SMGs are detected in the $H_{160}$-band imaging above a typical limiting  magnitude of 27.8 mag, yielding a detection rate of 79\,$\pm$\,17\%. Most (80\%) of the non-detections are sources with $S_{870} < $\,3\,mJy, and about half of the $S_{870} < $\,3\,mJy SMGs are undetected in our $H_{160}$-band imaging. In addition, 61\,$\pm$\,16\% of the 38 $H_{160}$-detected ALESS SMGs have more than one $H_{160}$-band component based on our S\'{e}rsic profile fitting using {\sc galfit}. {In contrast}, only 10\% of the comparison sample of lower-luminosity field galaxies have multiple $H_{160}$-band components.

\item We visually classified the $H_{160}$-band morphologies of these galaxies and found that 82\,$\pm$\,9\% of the $z=$\,1--3 ALESS SMGs appear to have disturbed morphologies, meaning they are either irregular or interacting systems. In comparison, 48\,$\pm$\,6\% of the lower-luminosity field sample at similar redshifts appear to be disturbed. 

\item Using {\sc galfit} to derive S\'{e}rsic profile fits on the $H_{160}$-band imaging, the SMGs at $z = $\,1--3 have a median S\'{e}rsic index of $n = $\,1.2\,$\pm$\,0.3 and a median effective radius (half-light radius) of $r_e = $\,4.4$^{+1.1}_{-0.5}$. We did not find any correlation between the S\'{e}rsic index, effective radius, $L_{\rm IR}$, redshift, or $H_{160}$. 

\item {Since SMGs are likely to be short-lived with an expected lifetime of $\sim$\,100\,Myr, along with evidence of fast quenching SFHs ($\tau < $\,250\,Myr) for $z \sim $\,2 quiescent galaxies,  SMGs at $z=$\,2--3 have many of the properties expected for the progenitors of the $z \sim $\,2 quiescent galaxies. Given the claimed fast evolution timescale, we  argue that major mergers are the main mechanism that drive the transformation of the stellar distribution, in both effective radius (from $r_e \sim $\,4\,kpc to $r_e =$\,1--2\,kpc) and S\'{e}rsic index (from $n \sim $\,1 to $n \sim$\,4), between $z=$\,2--3 SMGs and $z\sim$\,2 quiescent galaxies. Specifically, our findings that a high fraction of SMGs have a disturbed morphology, and significant offsets between the dusty star-forming regions and the stellar distributions, suggest that the majority of SMGs at $z=$\,2--3 are early/mid-stage mergers.}

\end{enumerate}

\vspace{5mm}
\section*{Acknowledgments}
{We thank the referee for a very thorough report that improves the manuscript.} We thank Jacob Head for useful discussions on {\sc galfit} and Anton Koekemoer for helpful suggestions on the usage of CANDELS maps. This research made use of Astropy, a community-developed core Python package for Astronomy \citep{Astropy-Collaboration:2013aa}. This research has made use of NASA's Astrophysics Data System. C.-C.C., I.R.S. acknowledge support from the ERC Advanced Investigator programme DUSTYGAL 321334. I.R.S. also acknowledges support from a Royal Society/Wolfson Merit Award and STFC through grant number ST/L00075X/1. Support for program \#12866 was provided by NASA through a grant from the Space Telescope Science Institute, which is operated by the Association of Universities for Research in Astronomy, Inc., under NASA contract NAS 5-26555. This paper makes use of the following ALMA data: ADS/JAO.ALMA\#2011.1.00294.S. ALMA is a partnership of ESO (representing its member states), NSF (USA) and NINS (Japan), together with NRC (Canada) and NSC and ASIAA (Taiwan), in cooperation with the Republic of Chile. The Joint ALMA Observatory is operated by ESO, AUI/NRAO and NAOJ.

\bibliography{bib}

\appendix
\section{Tables and images}

\clearpage
\LongTables
\setlength{\tabcolsep}{0.065in}
\begin{deluxetable*}{lcccccrrcccc}
\tablewidth{10in}
\tablecolumns{12}
\tablecaption{GALFIT Results on MAIN ALESS SMGs\label{tab:galfit}}
\tablehead{
\colhead{} & \colhead{R.A.} \vspace{-0.15cm}& \colhead{Dec.} & \colhead{} & \colhead{$H_{160}$} & \colhead{} & \colhead{$\Delta$R.A.} & \colhead{$\Delta$Dec.} & \colhead{$H_{160g}$} & \colhead{$r_e$} & \colhead{} & \colhead{} \\ 
 \colhead{ALESS ID} \vspace{-0.15cm}& \colhead{} & \colhead{} & \colhead{$z_{photo}$} & \colhead{} & \colhead{HID} & \colhead{} & \colhead{} & \colhead{} & \colhead{} & \colhead{$n$} & \colhead{$\chi^2_\nu$} \\
 \colhead{} & \colhead{[Degree]} & \colhead{[Degree]} & \colhead{} & \colhead{[ABmag]} & \colhead{} & \colhead{[$''$]} & \colhead{[$''$]} & \colhead{[ABmag]} & \colhead{[kpc]} & \colhead{} & \colhead{}  \\
 \colhead{(1)} & \colhead{(2)} & \colhead{(3)} & \colhead{(4)} & \colhead{(5)} & \colhead{(6)} & \colhead{(7)} & \colhead{(8)} & \colhead{(9)} & \colhead{(10)} & \colhead{(11)} & \colhead{(12)}  }
\startdata
ALESS 001.1 & 53.31027 & $-$27.93737 & 4.34$^{+2.66}_{-1.43}$ & $>$27.8 & \ldots & \ldots & \ldots & \ldots & \ldots & \ldots & \ldots\\
ALESS 001.2 & 53.31006 & $-$27.93656 & 4.64$^{+2.34}_{-1.02}$ & 26.0$\pm$0.7 & H1 & $-$0.72 & 0.13 & 26.50$\pm$0.24 & 1.87$^{+0.61}_{-0.68}$ & 0.96$\pm$1.41 & 1.09\\
ALESS 001.3 & 53.30907 & $-$27.93676 & 2.84$^{+0.20}_{-0.30}$ & 26.0$\pm$0.7 & H1 & $-$0.25 & 0.01 & 25.69$\pm$0.29 & 3.34$^{+1.89}_{-1.89}$ & 3.73$\pm$2.59 & 0.68\\
ALESS 002.1 & 53.26119 & $-$27.94521 & 1.96$^{+0.27}_{-0.20}$ & 24.0$\pm$0.3 & H1 & 0.18 & $-$0.07 & 23.98$\pm$0.02 & 1.49$^{+0.05}_{-0.05}$ & 1.33$\pm$0.13 & 1.01\\
ALESS 002.2 & 53.26280 & $-$27.94525 & \ldots & 26.4$\pm$0.9 & H1 & $-$0.20 & $-$0.19 & 26.25$\pm$0.28 & \ldots & 1.56$\pm$1.29 & 0.81\\
ALESS 003.1 & 53.33960 & $-$27.92230 & 3.90$^{+0.50}_{-0.59}$ & 24.7$\pm$0.6 & H1 & 0.25 & $-$0.48 & 24.67$\pm$0.07 & 5.51$^{+0.68}_{-0.65}$ & 1.00 & 1.06\\
ALESS 005.1 & 52.87047 & $-$27.98584 & 2.86$^{+0.05}_{-0.04}$ & 25.6$\pm$0.7 & \ldots & \ldots & \ldots & \ldots & \ldots & \ldots & \ldots\\
ALESS 009.1 & 53.04724 & $-$27.86998 & 4.50$^{+0.54}_{-2.33}$ & 25.5$\pm$0.6 & H1 & 0.19 & $-$0.80 & 25.06$\pm$0.30 & 5.82$^{+2.77}_{-2.38}$ & 1.29$\pm$0.77 & 0.97\\
ALESS 010.1 & 53.07942 & $-$27.87078 & 2.02$^{+0.09}_{-0.09}$ & 24.0$\pm$0.4 & H1 & $-$0.26 & $-$0.25 & 24.82$\pm$0.02 & 1.18$^{+0.05}_{-0.05}$ & 1.99$\pm$0.33 & 0.88\\
 &  &  &  &  & H2 & $-$0.33 & 0.60 & 24.26$\pm$0.03 & 2.69$^{+0.12}_{-0.12}$ & 1.67$\pm$0.14 & 0.88\\
ALESS 013.1 & 53.20413 & $-$27.71439 & 3.25$^{+0.64}_{-0.46}$ & 24.8$\pm$0.4 & H1 & $-$1.52 & $-$0.24 & 24.54$\pm$0.56 & 2.10$^{+2.14}_{-2.14}$ & 3.88$\pm$3.85 & 1.28\\
ALESS 015.1 & 53.38903 & $-$27.99155 & 1.92$^{+0.62}_{-0.33}$ & 23.8$\pm$0.5 & H1 & 0.18 & $-$0.02 & 26.04$\pm$0.11 & 1.95$^{+0.18}_{-0.20}$ & 0.33$\pm$0.15 & 0.84\\
 &  &  &  &  & H2 & $-$0.41 & $-$0.18 & 24.47$\pm$0.09 & 7.62$^{+0.89}_{-0.94}$ & 1.06$\pm$0.16 & 0.84\\
 &  &  &  &  & H3 & 0.90 & $-$0.09 & 25.97$\pm$0.06 & 2.38$^{+0.19}_{-0.22}$ & 0.25$\pm$0.15 & 0.84\\
 &  &  &  &  & H4 & 0.54 & $-$0.96 & 26.16$\pm$0.08 & 3.35$^{+0.30}_{-0.32}$ & 0.20$\pm$0.13 & 0.84\\
ALESS 015.3 & 53.38998 & $-$27.99318 & \ldots & $>$27.8 & \ldots & \ldots & \ldots & \ldots & \ldots & \ldots & \ldots\\
ALESS 017.1$^a$ & 53.03041 & $-$27.85576 & 1.50$^{+0.10}_{-0.07}$ & 21.4$\pm$0.1 & H1 & 0.42 & 0.19 & 22.04$\pm$0.04 & 5.48$^{+0.10}_{-0.10}$ & 0.28$\pm$0.01 & 3.00\\
 &  &  &  &  & H2 & 0.43 & 0.27 & 22.10$\pm$0.04 & 1.28$^{+0.05}_{-0.05}$ & 1.51$\pm$0.11 & 3.00\\
ALESS 018.1 & 53.02034 & $-$27.77993 & 2.04$^{+0.10}_{-0.06}$ & 22.2$\pm$0.2 & H1 & $-$0.08 & 0.12 & 22.06$\pm$0.01 & 6.25$^{+0.04}_{-0.05}$ & 0.32$\pm$0.01 & 2.98\\
 &  &  &  &  & H2 & $-$1.60 & 0.29 & 26.19$\pm$0.07 & 1.11$^{+0.26}_{-0.26}$ & 0.25$\pm$1.08 & 2.98\\
ALESS 029.1 & 53.40375 & $-$27.96926 & 2.66$^{+2.94}_{-0.76}$ & 24.4$\pm$0.5 & H1 & 0.24 & 0.38 & 25.50$\pm$0.06 & 2.88$^{+0.27}_{-0.76}$ & 0.42$\pm$0.13 & 0.84\\
 &  &  &  &  & H2 & $-$0.38 & $-$0.47 & 24.72$\pm$0.33 & 6.79$^{+3.47}_{-3.85}$ & 2.79$\pm$1.12 & 0.84\\
ALESS 039.1 & 52.93763 & $-$27.57687 & 2.44$^{+0.18}_{-0.23}$ & 23.5$\pm$0.2 & H1 & 0.03 & $-$0.08 & 24.78$\pm$0.08 & 2.42$^{+0.09}_{-0.08}$ & 0.10$\pm$0.06 & 0.91\\
 &  &  &  &  & H2 & $-$0.49 & 0.05 & 23.33$\pm$0.34 & 12.09$^{+5.41}_{-5.40}$ & 2.10$\pm$0.67 & 0.91\\
ALESS 043.1 & 53.27767 & $-$27.80068 & 1.70$^{+0.20}_{-0.12}$ & 23.3$\pm$0.3 & H1 & 0.01 & 0.05 & 23.40$\pm$0.03 & 8.60$^{+0.35}_{-0.35}$ & 0.94$\pm$0.06 & 1.29\\
 &  &  &  &  & H2 & 0.21 & $-$0.94 & 25.11$\pm$0.42 & 6.71$^{+3.68}_{-3.68}$ & 1.68$\pm$0.91 & 1.29\\
ALESS 045.1$^a$ & 53.10526 & $-$27.87515 & 2.34$^{+0.26}_{-0.67}$ & 23.6$\pm$0.3 & H1 & $-$0.11 & $-$0.06 & 23.89$\pm$0.09 & 6.15$^{+0.77}_{-0.76}$ & 4.00 & 1.02\\
 &  &  &  &  & H2 & 0.14 & $-$0.36 & 24.83$\pm$0.04 & 3.27$^{+0.19}_{-0.17}$ & 0.52$\pm$0.10 & 1.02\\
ALESS 049.1 & 52.85300 & $-$27.84641 & 2.76$^{+0.11}_{-0.14}$ & 23.1$\pm$0.2 & H1 & $-$0.11 & 0.40 & 24.76$\pm$0.09 & 5.25$^{+0.53}_{-0.53}$ & 1.00 & 1.67\\
 &  &  &  &  & H2 & $-$0.42 & 0.30 & 22.91$\pm$0.09 & 4.43$^{+0.80}_{-0.80}$ & 4.90$\pm$0.63 & 1.67\\
ALESS 049.2 & 52.85196 & $-$27.84391 & 1.46$^{+0.07}_{-0.10}$ & 23.1$\pm$0.2 & H1 & $-$0.05 & $-$0.20 & 23.60$\pm$0.06 & 3.16$^{+0.17}_{-0.17}$ & 1.24$\pm$0.10 & 1.51\\
 &  &  &  &  & H2 & 0.17 & 0.38 & 24.16$\pm$0.11 & 2.90$^{+0.19}_{-0.19}$ & 0.18$\pm$0.04 & 1.51\\
ALESS 051.1 & 52.93775 & $-$27.74092 & 1.22$^{+0.03}_{-0.06}$ & 19.6$\pm$0.1 & H1 & 0.54 & 0.16 & 21.53$\pm$0.03 & 3.93$^{+0.19}_{-0.20}$ & 4.00 & 21.06\\
 &  &  &  &  & H2 & 1.40 & 1.14 & 24.66$\pm$0.25 & 2.81$^{+0.88}_{-0.88}$ & 0.84$\pm$0.73 & 21.06\\
 &  &  &  &  & H3 & $-$1.45 & $-$1.86 & 19.70$\pm$0.01 & 4.86$^{+0.07}_{-0.08}$ & 1.88$\pm$0.03 & 21.06\\
ALESS 055.1 & 53.25924 & $-$27.67651 & 2.05$^{+0.15}_{-0.13}$ & 23.2$\pm$0.3 & H1 & 0.43 & $-$0.39 & 23.06$\pm$0.01 & 3.32$^{+0.06}_{-0.06}$ & 1.09$\pm$0.03 & 8.66\\
ALESS 055.2 & 53.25898 & $-$27.67815 & \ldots & $>$30.3 & \ldots & \ldots & \ldots & \ldots & \ldots & \ldots & \ldots\\
ALESS 057.1$^a$ & 52.96635 & $-$27.89085 & 2.95$^{+0.05}_{-0.09}$ & 23.3$\pm$0.2 & H1 & 0.14 & 0.09 & 23.53$\pm$0.04 & 2.21$^{+0.11}_{-0.10}$ & 1.75$\pm$0.13 & 1.34\\
 &  &  &  &  & H2 & 0.25 & $-$0.39 & 24.60$\pm$0.12 & 3.23$^{+0.46}_{-0.46}$ & 1.30$\pm$0.28 & 1.34\\
ALESS 067.1$^a$ & 53.17998 & $-$27.92065 & 2.13$^{+0.05}_{-0.09}$ & 21.7$\pm$0.2 & H1 & 0.10 & $-$0.26 & 22.14$\pm$0.03 & 9.55$^{+0.39}_{-0.39}$ & 1.42$\pm$0.06 & 1.45\\
 &  &  &  &  & H2 & $-$0.51 & $-$0.22 & 24.22$\pm$0.02 & 1.61$^{+0.04}_{-0.04}$ & 0.35$\pm$0.08 & 1.45\\
 &  &  &  &  & H3 & $-$0.11 & $-$0.68 & 24.42$\pm$0.02 & 1.13$^{+0.04}_{-0.04}$ & 0.40$\pm$0.20 & 1.45\\
 &  &  &  &  & H4 & 0.96 & $-$0.99 & 23.57$\pm$0.01 & 1.64$^{+0.03}_{-0.03}$ & 0.77$\pm$0.05 & 1.45\\
ALESS 069.1 & 52.89073 & $-$27.99234 & 2.34$^{+0.27}_{-0.44}$ & 24.1$\pm$0.3 & H1 & 0.06 & 0.06 & 24.00$\pm$0.02 & 4.13$^{+0.17}_{-0.16}$ & 0.54$\pm$0.05 & 1.00\\
ALESS 069.2 & 52.89223 & $-$27.99136 & \ldots & $>$27.9 & \ldots & \ldots & \ldots & \ldots & \ldots & \ldots & \ldots\\
ALESS 069.3 & 52.89152 & $-$27.99398 & \ldots & $>$27.9 & \ldots & \ldots & \ldots & \ldots & \ldots & \ldots & \ldots\\
ALESS 073.1$^a$ & 53.12205 & $-$27.93881 & 5.18$^{+0.43}_{-0.45}$ & 24.0$\pm$0.3 & H1 & 0.03 & 0.01 & 23.93$\pm$0.01 & $<$ 1.19 & \ldots & 2.02\\
ALESS 074.1 & 53.28811 & $-$27.80477 & 1.80$^{+0.13}_{-0.13}$ & 23.4$\pm$0.2 & H1 & 0.02 & 0.37 & 23.42$\pm$0.05 & 5.69$^{+0.41}_{-0.41}$ & 1.65$\pm$0.12 & 0.94\\
 &  &  &  &  & H2 & $-$0.19 & $-$0.37 & 25.05$\pm$0.06 & 2.42$^{+0.13}_{-0.13}$ & 0.39$\pm$0.09 & 0.94\\
ALESS 079.1 & 53.08806 & $-$27.94083 & 2.04$^{+0.63}_{-0.31}$ & 24.6$\pm$0.5 & H1 & $-$0.10 & $-$0.06 & 23.95$\pm$0.05 & 7.58$^{+0.58}_{-0.67}$ & 1.00 & 1.17\\
ALESS 079.2 & 53.09000 & $-$27.93999 & 1.55$^{+0.10}_{-0.18}$ & 21.9$\pm$0.1 & H1 & $-$0.22 & $-$0.24 & 23.20$\pm$0.03 & 3.15$^{+0.07}_{-0.08}$ & 0.30$\pm$0.03 & 2.11\\
 &  &  &  &  & H2 & $-$0.58 & $-$0.12 & 23.87$\pm$0.05 & 1.29$^{+0.05}_{-0.05}$ & 0.98$\pm$0.13 & 2.11\\
 &  &  &  &  & H3 & 0.11 & 0.63 & 22.32$\pm$0.06 & 8.09$^{+0.61}_{-0.61}$ & 1.88$\pm$0.13 & 2.11\\
 &  &  &  &  & H4 & $-$0.42 & $-$0.75 & 23.03$\pm$0.02 & 4.29$^{+0.08}_{-0.08}$ & 0.76$\pm$0.03 & 2.11\\
ALESS 079.4 & 53.08826 & $-$27.94181 & \ldots & $>$28.7 & \ldots & \ldots & \ldots & \ldots & \ldots & \ldots & \ldots\\
ALESS 082.1 & 53.22499 & $-$27.63747 & 2.10$^{+3.27}_{-0.44}$ & 24.0$\pm$0.4 & H1 & $-$0.13 & $-$0.23 & 24.14$\pm$0.03 & 4.19$^{+0.22}_{-1.14}$ & 0.71$\pm$0.07 & 1.18\\
 &  &  &  &  & H2 & $-$1.25 & 0.38 & 25.88$\pm$0.13 & 2.55$^{+0.44}_{-0.81}$ & 1.21$\pm$0.64 & 1.18\\
ALESS 087.1 & 53.21202 & $-$27.52819 & 3.20$^{+0.08}_{-0.47}$ & 22.5$\pm$0.1 & H1 & $-$0.24 & $-$0.07 & 23.92$\pm$0.02 & 1.03$^{+0.06}_{-0.03}$ & 0.49$\pm$0.08 & 1.95\\
 &  &  &  &  & H2 & $-$0.74 & $-$0.30 & 21.97$\pm$0.13 & 12.92$^{+2.70}_{-2.64}$ & 3.88$\pm$0.43 & 1.95\\
ALESS 087.3 & 53.21361 & $-$27.53075 & \ldots & $>$27.9 & \ldots & \ldots & \ldots & \ldots & \ldots & \ldots & \ldots\\
ALESS 088.1 & 52.97818 & $-$27.89486 & 1.84$^{+0.12}_{-0.11}$ & 22.3$\pm$0.3 & H1 & 0.04 & 0.32 & 24.48$\pm$0.05 & 6.63$^{+0.48}_{-0.48}$ & 1.00 & 1.26\\
 &  &  &  &  & H2 & $-$0.38 & $-$0.34 & 23.57$\pm$0.01 & 2.19$^{+0.04}_{-0.04}$ & 0.86$\pm$0.06 & 1.26\\
 &  &  &  &  & H3 & $-$0.41 & 0.48 & 24.72$\pm$0.03 & 4.60$^{+0.16}_{-0.16}$ & 0.08$\pm$0.07 & 1.26\\
 &  &  &  &  & H4 & 0.75 & 0.13 & 23.77$\pm$0.11 & 6.56$^{+0.92}_{-0.92}$ & 4.00 & 1.26\\
 &  &  &  &  & H5 & 1.57 & $-$0.06 & 23.47$\pm$0.25 & 16.20$^{+3.82}_{-3.82}$ & 4.00 & 1.26\\
 &  &  &  &  & H6 & 1.61 & 0.30 & 24.57$\pm$0.09 & 1.85$^{+0.18}_{-0.18}$ & 1.57$\pm$0.36 & 1.26\\
ALESS 088.11 & 52.97895 & $-$27.89378 & 2.56$^{+0.04}_{-0.12}$ & 23.2$\pm$0.2 & H1 & $-$0.10 & 0.64 & 23.28$\pm$0.04 & 1.65$^{+0.12}_{-0.12}$ & 4.03$\pm$0.43 & 1.22\\
 &  &  &  &  & H2 & 0.06 & 1.12 & 25.78$\pm$0.40 & 3.29$^{+1.91}_{-1.91}$ & 4.00 & 1.22\\
 &  &  &  &  & H3 & $-$0.26 & 1.36 & 25.11$\pm$0.14 & 6.08$^{+1.12}_{-1.11}$ & 1.00 & 1.22\\
ALESS 088.2 & 52.98080 & $-$27.89453 & \ldots & $>$27.9 & \ldots & \ldots & \ldots & \ldots & \ldots & \ldots & \ldots\\
ALESS 088.5 & 52.98252 & $-$27.89645 & 2.30$^{+0.11}_{-0.50}$ & 23.1$\pm$0.2 & H1 & 0.19 & 0.38 & 25.79$\pm$0.08 & 1.85$^{+0.23}_{-0.22}$ & 0.57$\pm$0.40 & 1.49\\
 &  &  &  &  & H2 & $-$0.28 & $-$0.41 & 23.67$\pm$0.05 & 4.89$^{+0.34}_{-0.32}$ & 1.06$\pm$0.10 & 1.49\\
 &  &  &  &  & H3 & $-$0.49 & $-$0.95 & 24.04$\pm$0.04 & 1.84$^{+0.07}_{-0.06}$ & 0.55$\pm$0.06 & 1.49\\
ALESS 092.2 & 52.90891 & $-$27.72872 & 1.90$^{+0.27}_{-0.75}$ & 24.7$\pm$0.6 & H1 & 0.48 & $-$0.40 & 25.85$\pm$0.05 & 1.61$^{+0.00}_{-0.14}$ & 0.24$\pm$0.16 & 0.96\\
 &  &  &  &  & H2 & $-$0.21 & 0.98 & 24.74$\pm$0.23 & 3.59$^{+0.00}_{-1.17}$ & 1.98$\pm$0.79 & 0.96\\
ALESS 094.1 & 53.28164 & $-$27.96828 & 2.87$^{+0.37}_{-0.64}$ & 24.1$\pm$0.3 & H1 & 0.26 & $-$0.40 & 23.67$\pm$0.11 & 7.92$^{+1.07}_{-1.02}$ & 1.00 & 1.88\\
ALESS 112.1 & 53.20360 & $-$27.52036 & 1.95$^{+0.15}_{-0.26}$ & 22.2$\pm$0.2 & H1 & $-$0.05 & 0.15 & 24.16$\pm$0.12 & 0.87$^{+0.03}_{-0.03}$ & 1.08$\pm$0.21 & 1.31\\
 &  &  &  &  & H2 & $-$0.18 & $-$0.02 & 23.81$\pm$0.05 & 3.46$^{+0.06}_{-0.06}$ & 0.16$\pm$0.02 & 1.31\\
 &  &  &  &  & H3 & 0.08 & 0.17 & 23.03$\pm$0.27 & 7.32$^{+4.46}_{-4.46}$ & 4.58$\pm$1.47 & 1.31\\
 &  &  &  &  & H4 & $-$0.59 & 0.37 & 23.48$\pm$0.31 & 10.15$^{+4.08}_{-4.08}$ & 2.22$\pm$0.59 & 1.31\\
 &  &  &  &  & H5 & 0.00 & 1.07 & 25.55$\pm$0.05 & 1.88$^{+0.15}_{-0.15}$ & 0.25$\pm$0.20 & 1.31\\
ALESS 114.1 & 52.96036 & $-$27.74593 & \ldots & $>$27.9 & \ldots & \ldots & \ldots & \ldots & \ldots & \ldots & \ldots\\
ALESS 114.2$^a$ & 52.96294 & $-$27.74369 & 1.56$^{+0.07}_{-0.07}$ & 21.1$\pm$0.1 & H1 & $-$0.13 & $-$0.09 & 21.32$\pm$0.02 & 3.12$^{+0.06}_{-0.06}$ & 1.65$\pm$0.08 & 2.48\\
 &  &  &  &  & H2 & $-$0.05 & $-$0.04 & 22.40$\pm$0.05 & 0.96$^{+0.01}_{-0.01}$ & 0.17$\pm$0.06 & 2.48\\
ALESS 116.1 & 52.97634 & $-$27.75804 & 3.54$^{+1.47}_{-0.87}$ & $>$28.0 & \ldots & \ldots & \ldots & \ldots & \ldots & \ldots & \ldots\\
ALESS 116.2 & 52.97683 & $-$27.75874 & 4.01$^{+1.19}_{-2.19}$ & 26.7$\pm$1.0 & H1 & 0.15 & $-$0.05 & 25.84$\pm$0.22 & 5.40$^{+1.91}_{-1.66}$ & 1.00 & 0.64\\
ALESS 118.1 & 52.84135 & $-$27.82816 & 2.26$^{+0.50}_{-0.23}$ & 24.3$\pm$0.3 & H1 & $-$0.14 & $-$0.16 & 24.64$\pm$0.11 & 2.06$^{+0.34}_{-0.34}$ & 2.10$\pm$0.56 & 1.38\\
\enddata
\tablenotetext{a}{X-ray identified Active Galactic Nuclei (AGN) in \citet{Wang:2013aa}}
\tablecomments{In Columns 1, 2, and 3 we give the source IDs and the sky positions in right ascension (R.A.) and declination (Dec.), both J2000, adopted from \citealt{Hodge:2013lr}. In Column 4, we give the photometric redshifts derived by \citet{Simpson:2014aa}, in which the errors show the 1\,$\sigma$ uncertainty. In Column 5 we show the measured total $H_{160}$-band flux densities in AB magnitude using {\sc SExtractor}, in which the sensitivity limit is given for the non-detections. Note the uncertainties of the $H_{160}$-band flux include Poisson errors based on our adopted CCD gain of 2.4 for WFC3/IR \citep{Dressel:2014aa}, and in our case the Poisson errors dominate over sky noise. In Column 6 we list the designated IDs for each component detected in our $H_{160}$-band imaging. In Columns 7 -- 12 we present the {\sc GALFIT} results. In Columns 7 and 8 we show the displacements in sky positions in arcsecond between the $H_{160}$-band component and the ALMA 870\,$\mu$m positions given in Columns 2 and 3, in which the positive values in R.A. and Decl. indicate displacements toward {east} and north with respect to the ALMA positions. The mean errors of displacements based on the model fit are $0\farcs012\pm0\farcs01$ in RA. and $0\farcs011\pm0\farcs009$ in Decl.. In Column 9 we give the $H_{160}$-band flux densities in AB magnitudes of the model S\'{e}rsic profile with the effective radius (r$_e$) and the S\'{e}rsic index ($n$) given in Columns 10 and 11. The errors of $r_e$ include the errors of both the photometric redshifts and the model fit, and the S\'{e}rsic indices without errors are fixed values, 1.0 or 4.0, in model fit {(last paragraph of \autoref{sec:galfit})}. The reduced chi-square ($\chi^2_\nu$) of the GALFIT S\'{e}rsic models are shown in Column 12. The $\chi^2_\nu$ values are the same for every component corresponding to the same SMG because {\sc GALFIT} performed simultaneous fitting to the masked pixels.}

\end{deluxetable*}
\clearpage

\setcounter{figure}{0}
\begin{figure*}
      \includegraphics[page=1]{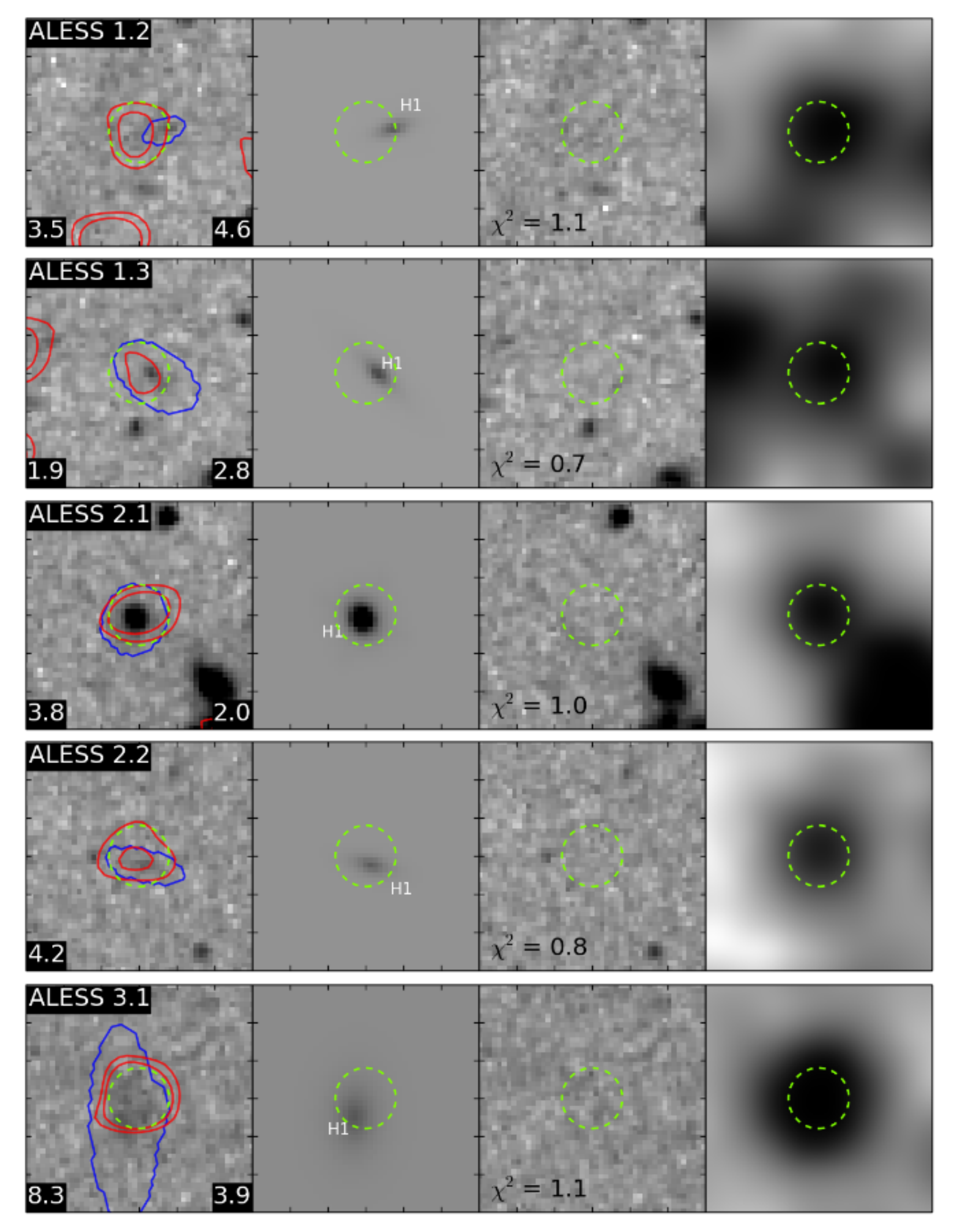}
       \caption{Postage stamps of $H$-band imaging on each MAIN ALESS SMGs. For each source, the left-to-right panels are the original $H$-band imaging, {\sc GALFIT} S\'{e}rsic models within the masked area (blue contours), residuals after subtracting the S\'{e}rsic models with the reduced $\chi^2$ noted, and the IRAC 3.6\,$\mu$m, respectively. The size of each box is $6''\times6''$, with the green circles representing the typical ALMA synthesized beam of 1$\farcs$6 FWHM. The ALMA contour at 3 and 5\,$\sigma$ is shown in red contours at the left panel for each source. The black numbers at the bottom-left are $S_{870}$, and those at the bottom-right corner, if any, are redshifts.}
    \label{postage}
\end{figure*}
\setcounter{figure}{0}
\begin{figure*}
      \includegraphics[page=2]{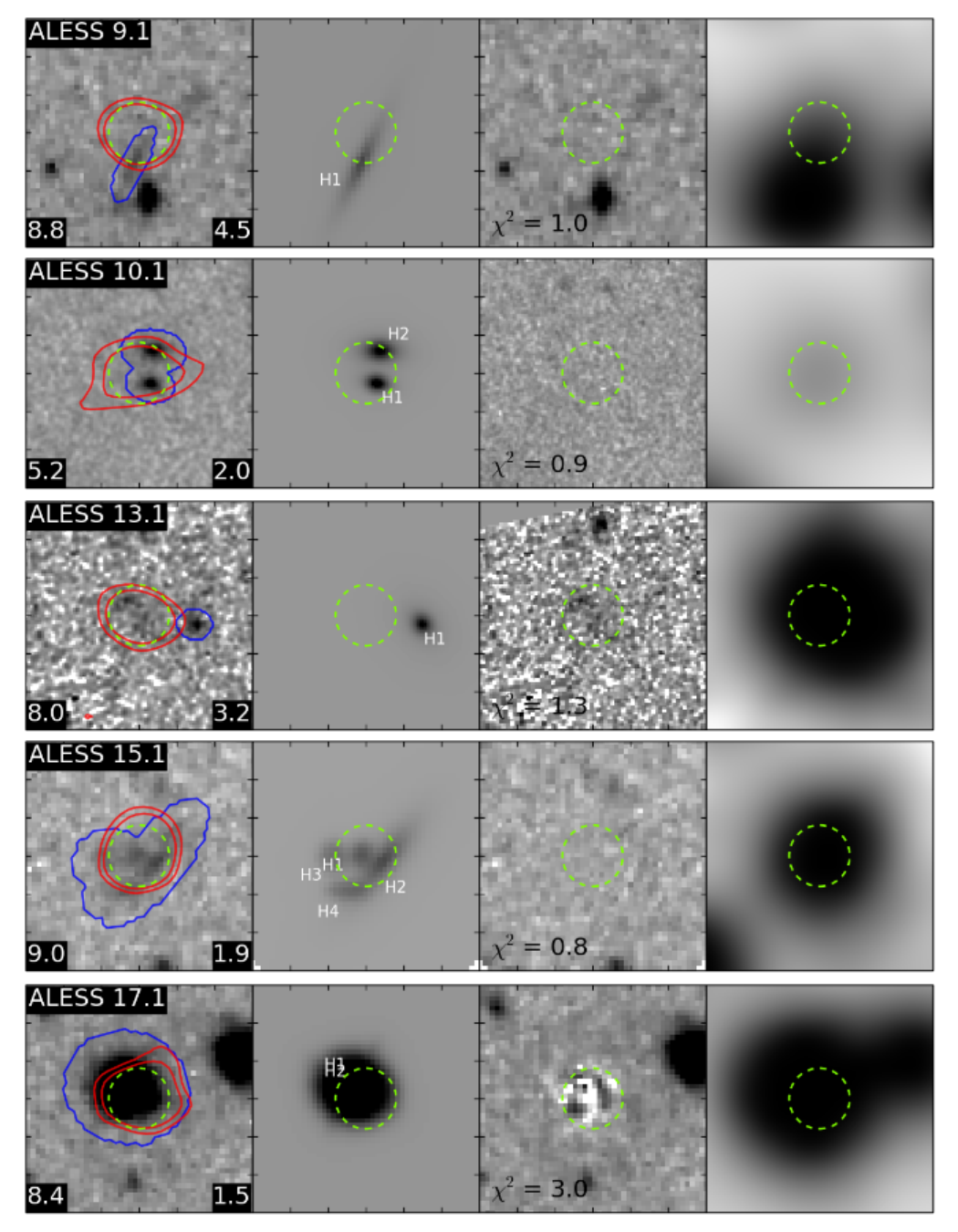}
       \caption{- {\it continued}}
\end{figure*}
\setcounter{figure}{0}
\begin{figure*}
      \includegraphics[page=3]{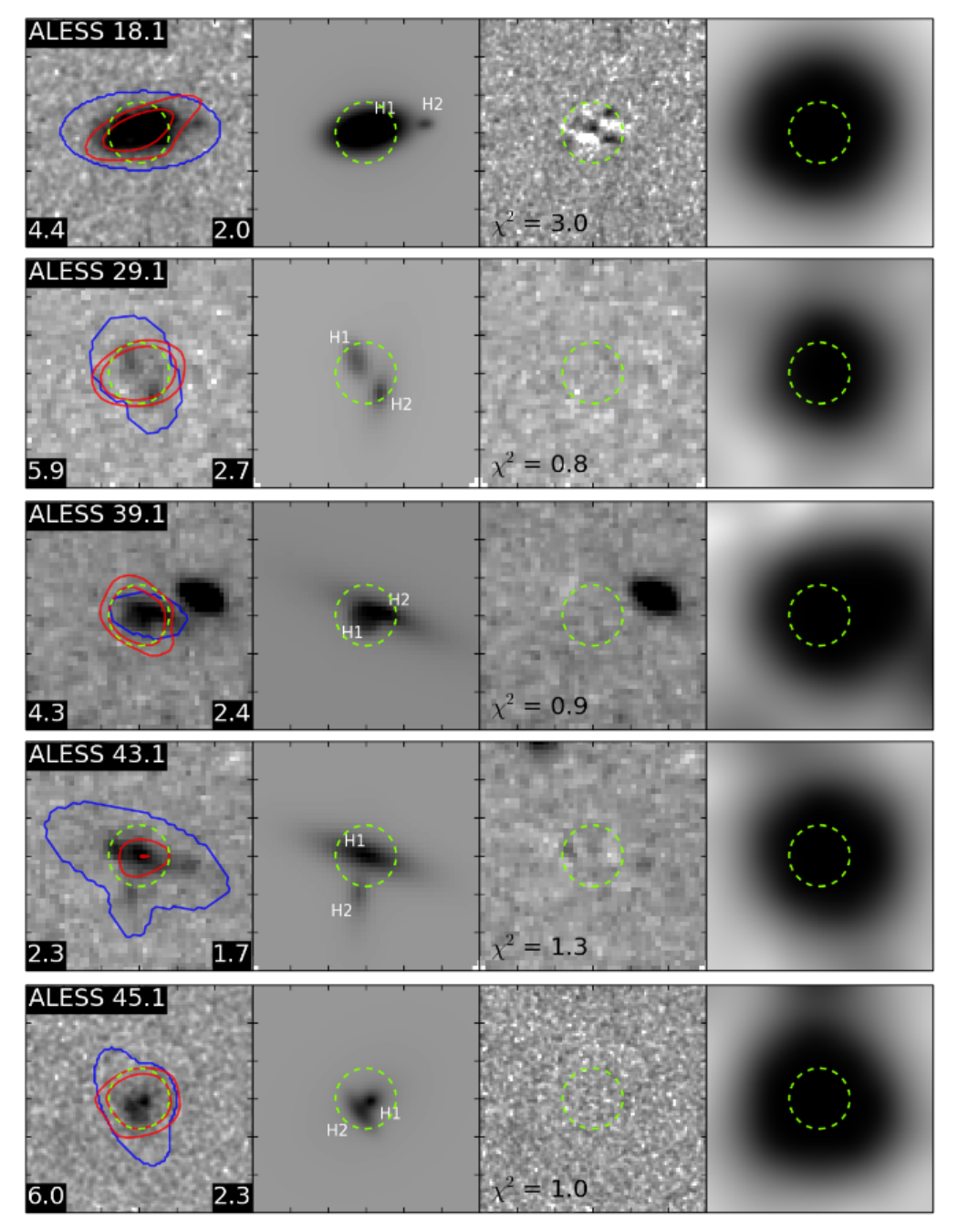}
       \caption{- {\it continued}}
\end{figure*}
\setcounter{figure}{0}
\begin{figure*}
      \includegraphics[page=4]{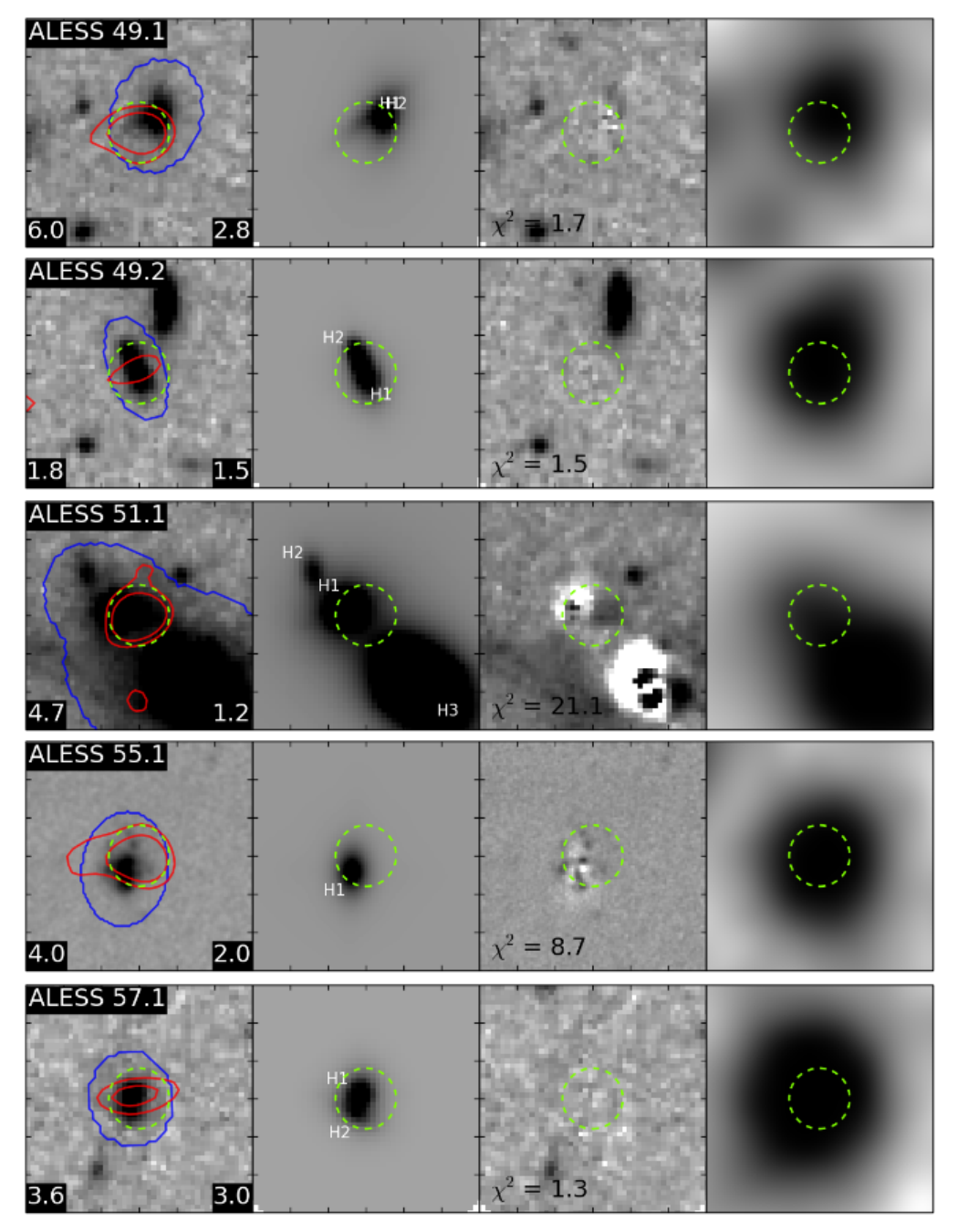}
       \caption{- {\it continued}}
\end{figure*}
\setcounter{figure}{0}
\begin{figure*}
      \includegraphics[page=5]{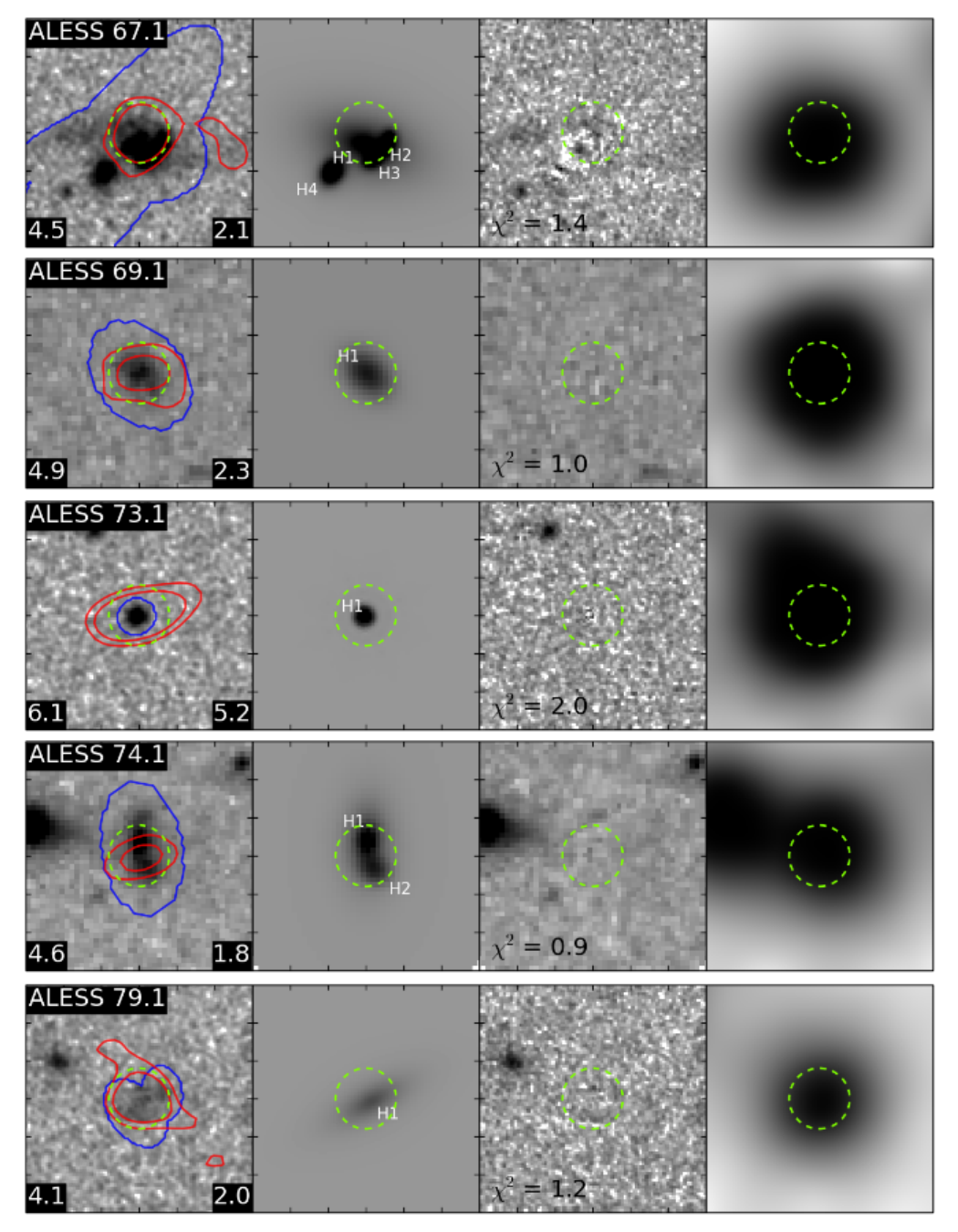}
       \caption{- {\it continued}}
\end{figure*}
\setcounter{figure}{0}
\begin{figure*}
      \includegraphics[page=6]{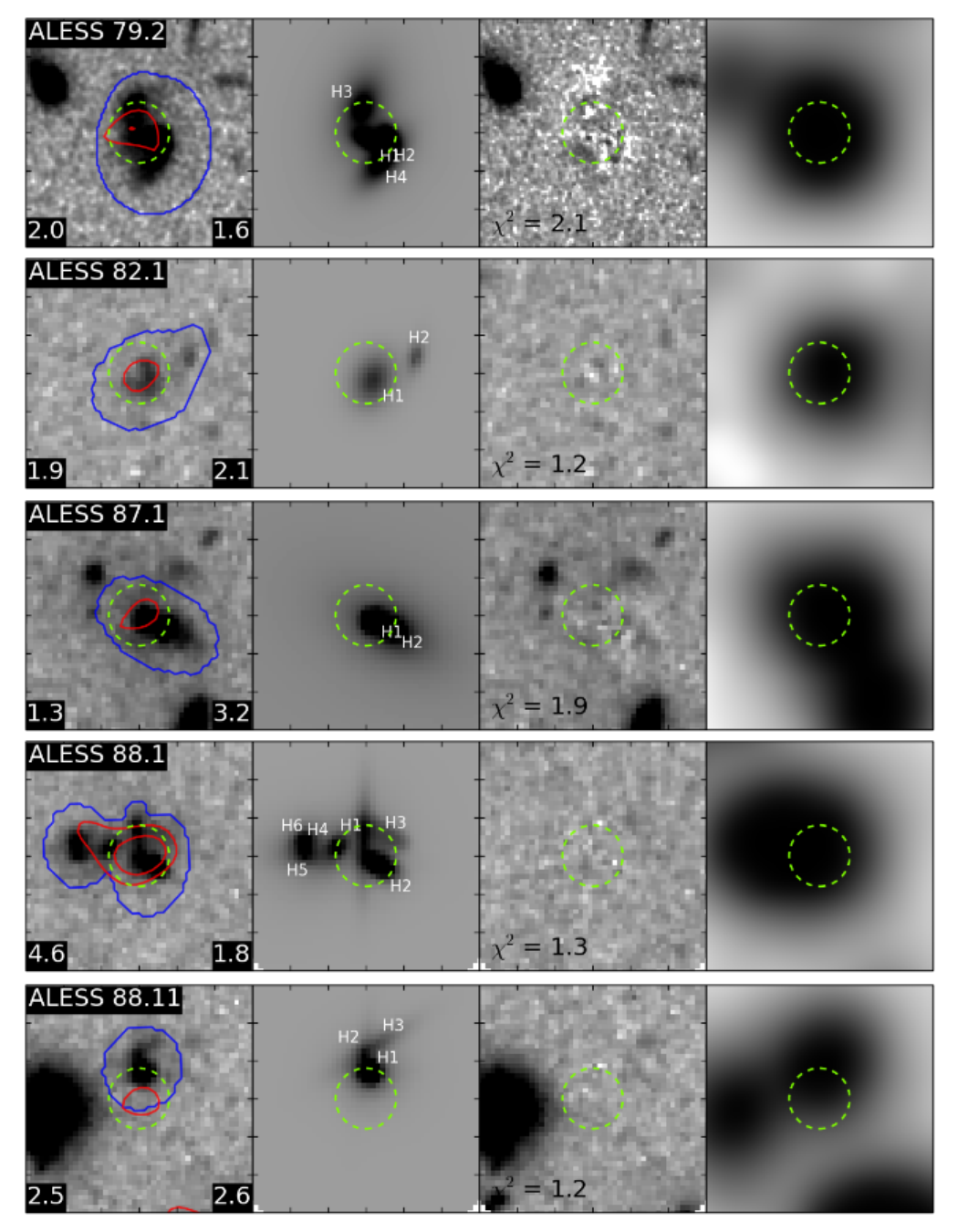}
       \caption{- {\it continued}}
\end{figure*}

\setcounter{figure}{0}
\begin{figure*}
      \includegraphics[page=7]{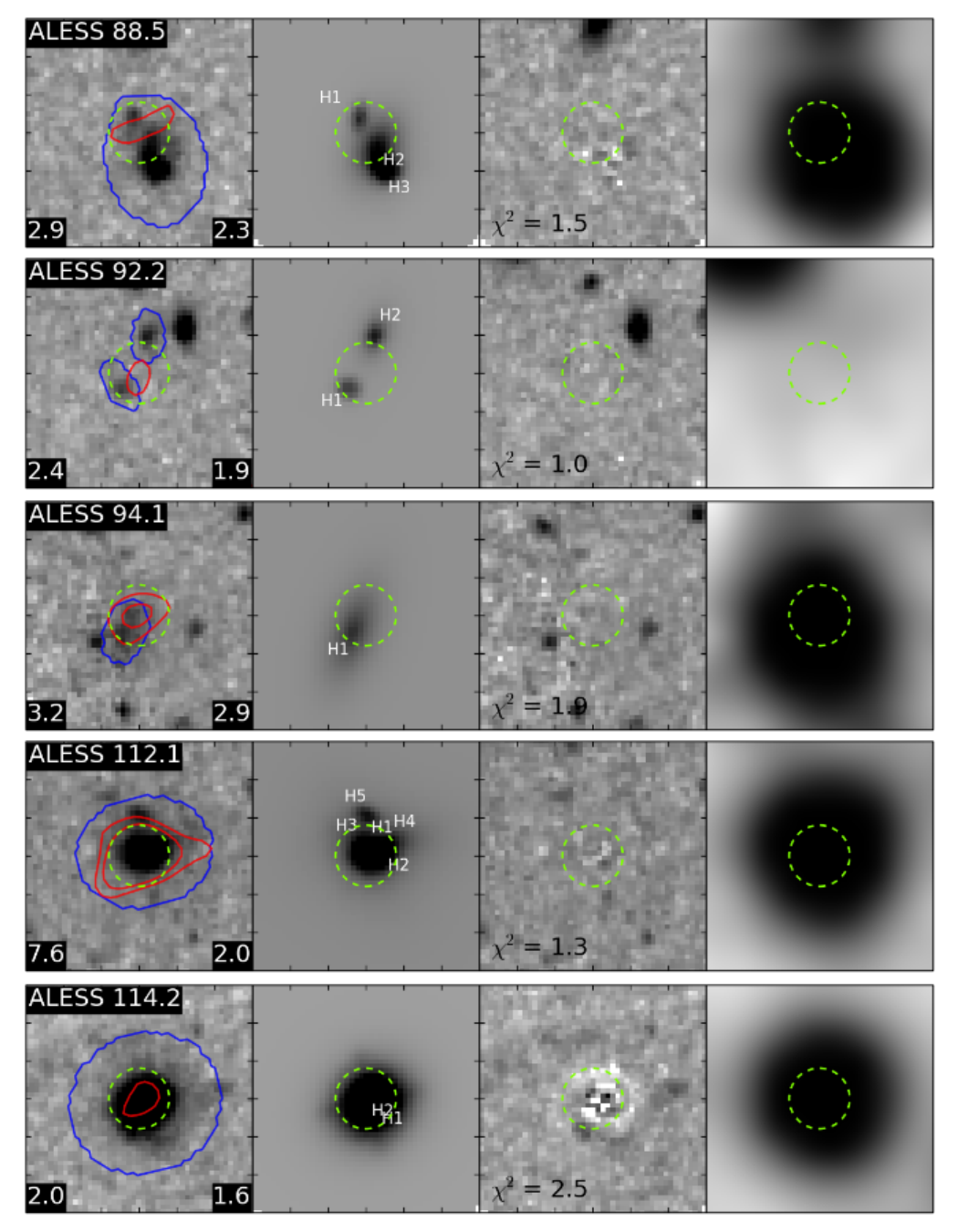}
       \caption{- {\it continued}}
\end{figure*}

\setcounter{figure}{0}
\begin{figure*}
      \includegraphics[page=8]{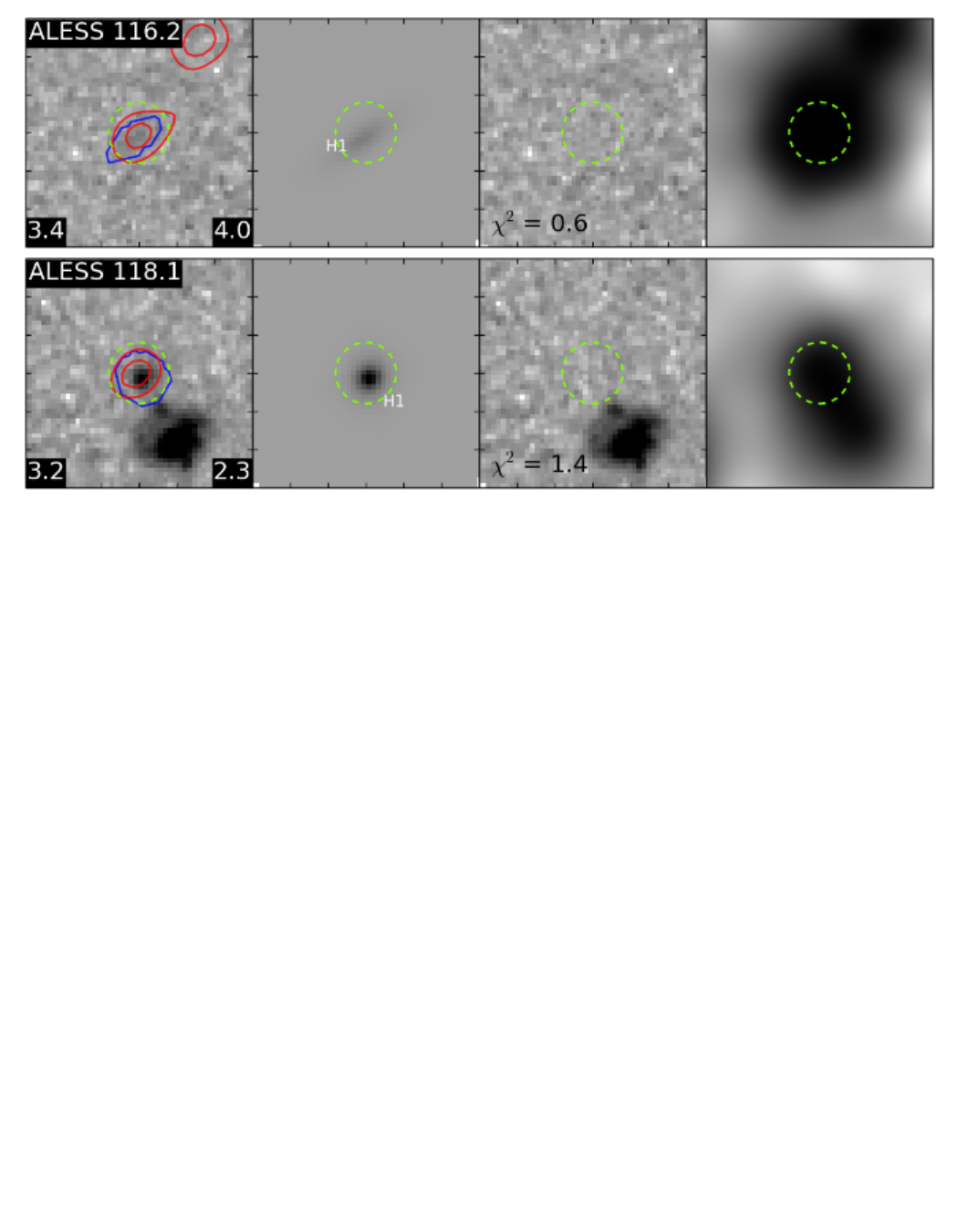}
       \caption{- {\it continued}}
\end{figure*}

\begin{figure*}
\centering
      \includegraphics[scale=1.0]{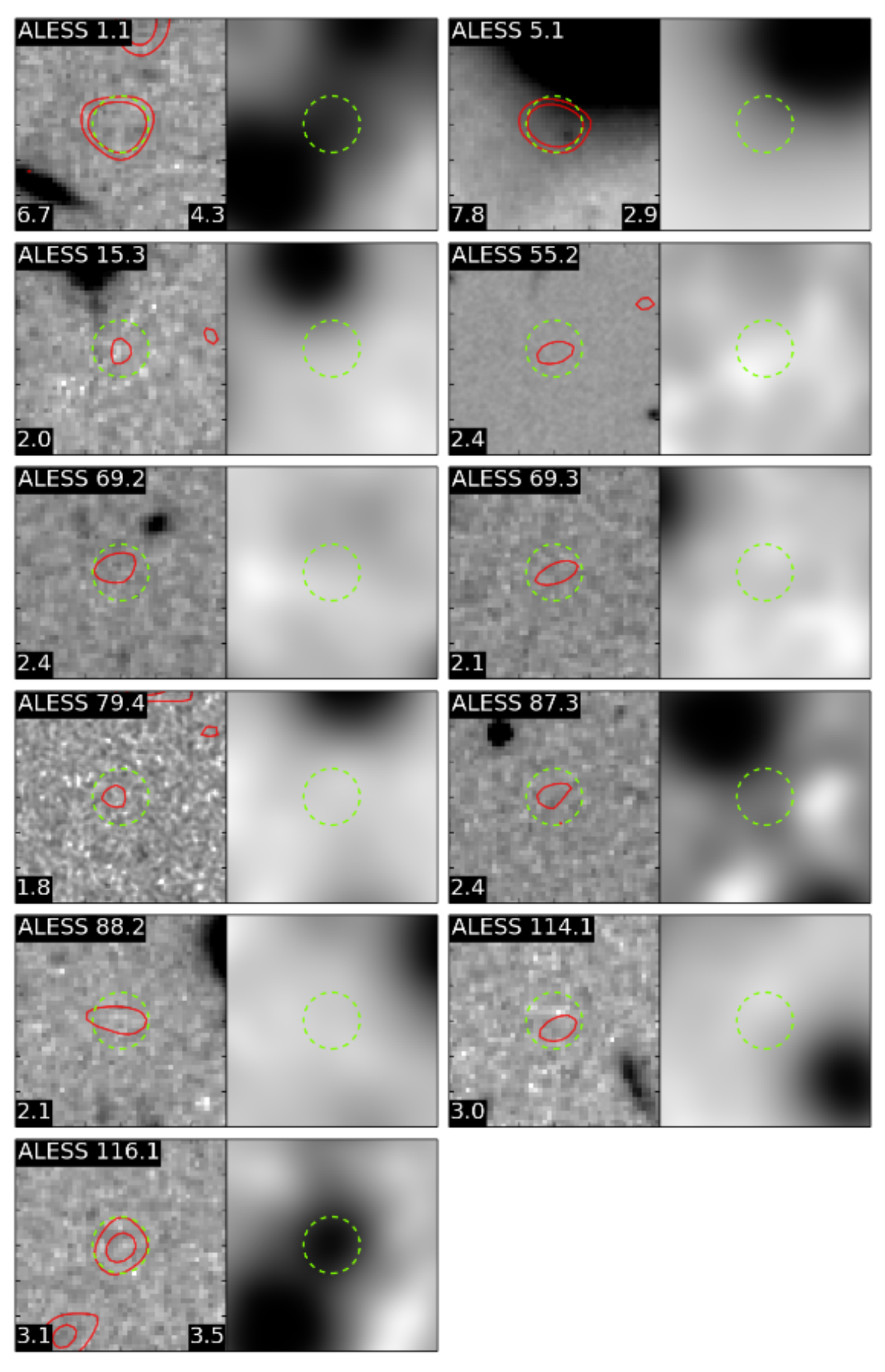}
       \caption{Postage stamps for non-detections of the MAIN ALESS SMGs except ALESS 5.1, which is detected but heavily contaminated by a nearby galaxy to conduct a S\'{e}rsic model fit. Same with \autoref{postage} but without the middle panels from {\sc GALFIT}.}
    \label{postage1}
\end{figure*}

\clearpage
\LongTables
\setlength{\tabcolsep}{0.065in}
\begin{deluxetable*}{lcccccrrcccc}
\tablewidth{0pc}
\tablecolumns{12}
\tablecaption{GALFIT Results on SUPPLEMENTARY ALESS SMGs\label{tab:galfitsup}}
\tablehead{
\colhead{} & \colhead{R.A.} \vspace{-0.15cm}& \colhead{Decl.} & \colhead{} & \colhead{$H_{160}$} & \colhead{} & \colhead{$\Delta$R.A.} & \colhead{$\Delta$Decl.} & \colhead{$H_{160g}$} & \colhead{R$_e$} & \colhead{} & \colhead{} \\ 
 \colhead{ALESS ID} \vspace{-0.15cm}& \colhead{} & \colhead{} & \colhead{$z_{photo}$} & \colhead{} & \colhead{HID} & \colhead{} & \colhead{} & \colhead{} & \colhead{} & \colhead{$n$} & \colhead{$\chi^2$} \\
 \colhead{} & \colhead{[Degree]} & \colhead{[Degree]} & \colhead{} & \colhead{[ABmag]} & \colhead{} & \colhead{[$''$]} & \colhead{[$''$]} & \colhead{[ABmag]} & \colhead{[kpc]} & \colhead{} & \colhead{}  \\
  \colhead{(1)} & \colhead{(2)} & \colhead{(3)} & \colhead{(4)} & \colhead{(5)} & \colhead{(6)} & \colhead{(7)} & \colhead{(8)} & \colhead{(9)} & \colhead{(10)} & \colhead{(11)} & \colhead{(12)}  }
\startdata
ALESS 003.2 & 53.34246 & $-$27.92249 & 1.39$^{+0.37}_{-0.34}$ & $>$27.8 & \ldots & \ldots & \ldots & \ldots & \ldots & \ldots & \ldots\\
ALESS 003.3 & 53.33629 & $-$27.92056 & \ldots & $>$27.9 & \ldots & \ldots & \ldots & \ldots & \ldots & \ldots & \ldots\\
ALESS 003.4 & 53.34164 & $-$27.91938 & \ldots & $>$27.9 & \ldots & \ldots & \ldots & \ldots & \ldots & \ldots & \ldots\\
ALESS 015.2 & 53.39188 & $-$27.99172 & 0.72$^{+0.65}_{-0.72}$ & 25.6$\pm$0.9 & H1 & $-$0.40 & 0.83 & 26.56$\pm$0.09 & 1.43$^{+0.29}_{-1.44}$ & 0.24$\pm$0.46 & 0.91\\
 &  &  &  &  & H2 & $-$0.88 & 1.40 & 26.02$\pm$0.08 & 2.55$^{+0.51}_{-2.57}$ & 0.35$\pm$0.16 & 0.91\\
ALESS 015.6 & 53.38819 & $-$27.99505 & \ldots & $>$27.9 & \ldots & \ldots & \ldots & \ldots & \ldots & \ldots & \ldots\\
ALESS 017.2 & 53.03444 & $-$27.85547 & 2.10$^{+0.62}_{-1.35}$ & 25.1$\pm$0.7 & H1 & $-$0.41 & $-$0.48 & 25.31$\pm$0.08 & 4.04$^{+0.00}_{-0.79}$ & 1.00 & 0.93\\
ALESS 017.3 & 53.03072 & $-$27.85942 & 2.58$^{+0.14}_{-0.25}$ & 24.3$\pm$0.5 & H1 & 0.15 & 0.72 & 25.90$\pm$0.05 & 1.13$^{+0.11}_{-0.11}$ & 0.64$\pm$0.35 & 0.71\\
 &  &  &  &  & H2 & $-$0.53 & $-$0.66 & 24.50$\pm$0.02 & 1.42$^{+0.05}_{-0.05}$ & 0.87$\pm$0.12 & 0.71\\
ALESS 034.1 & 53.07483 & $-$27.87591 & 1.86$^{+0.29}_{-0.32}$ & 26.1$\pm$0.8 & H1 & $-$0.23 & $-$0.49 & 26.48$\pm$0.04 & 0.81$^{+0.12}_{-0.12}$ & 1.00 & 0.63\\
ALESS 038.1 & 53.29515 & $-$27.94450 & 2.47$^{+0.10}_{-0.05}$ & 22.2$\pm$0.2 & H1 & $-$0.34 & $-$0.37 & 24.00$\pm$0.03 & 3.51$^{+0.15}_{-0.16}$ & 0.89$\pm$0.07 & 0.99\\
 &  &  &  &  & H2 & 0.95 & 0.20 & 22.37$\pm$0.01 & 3.05$^{+0.03}_{-0.04}$ & 0.99$\pm$0.02 & 0.99\\
ALESS 039.2 & 52.93567 & $-$27.57865 & 0.79$^{+0.28}_{-0.17}$ & $>$27.8 & \ldots & \ldots & \ldots & \ldots & \ldots & \ldots & \ldots\\
ALESS 043.3 & 53.27612 & $-$27.79853 & 1.98$^{+0.88}_{-0.99}$ & $>$27.8 & \ldots & \ldots & \ldots & \ldots & \ldots & \ldots & \ldots\\
ALESS 101.1 & 52.96499 & $-$27.76472 & 3.49$^{+3.52}_{-0.88}$ & 23.4$\pm$0.4 & H1 & $-$0.10 & $-$0.19 & 24.50$\pm$0.22 & 9.39$^{+2.18}_{-3.34}$ & 1.08$\pm$0.22 & 1.14\\
 &  &  &  &  & H2 & 0.66 & $-$0.91 & 24.90$\pm$0.28 & 4.50$^{+2.19}_{-2.50}$ & 3.61$\pm$1.35 & 1.14\\
 &  &  &  &  & H3 & $-$1.14 & $-$0.84 & 24.99$\pm$0.03 & 1.72$^{+0.17}_{-0.50}$ & 0.52$\pm$0.09 & 1.14\\
 &  &  &  &  & H4 & $-$2.02 & $-$1.00 & 25.26$\pm$0.30 & 7.08$^{+2.97}_{-3.53}$ & 4.00 & 1.14\\
 &  &  &  &  & H5 & $-$1.80 & $-$1.63 & 26.26$\pm$0.09 & 1.38$^{+0.27}_{-0.46}$ & 0.80$\pm$0.60 & 1.14\\
\enddata
\end{deluxetable*}

\setcounter{figure}{2}
\begin{figure*}
      \includegraphics[page=1]{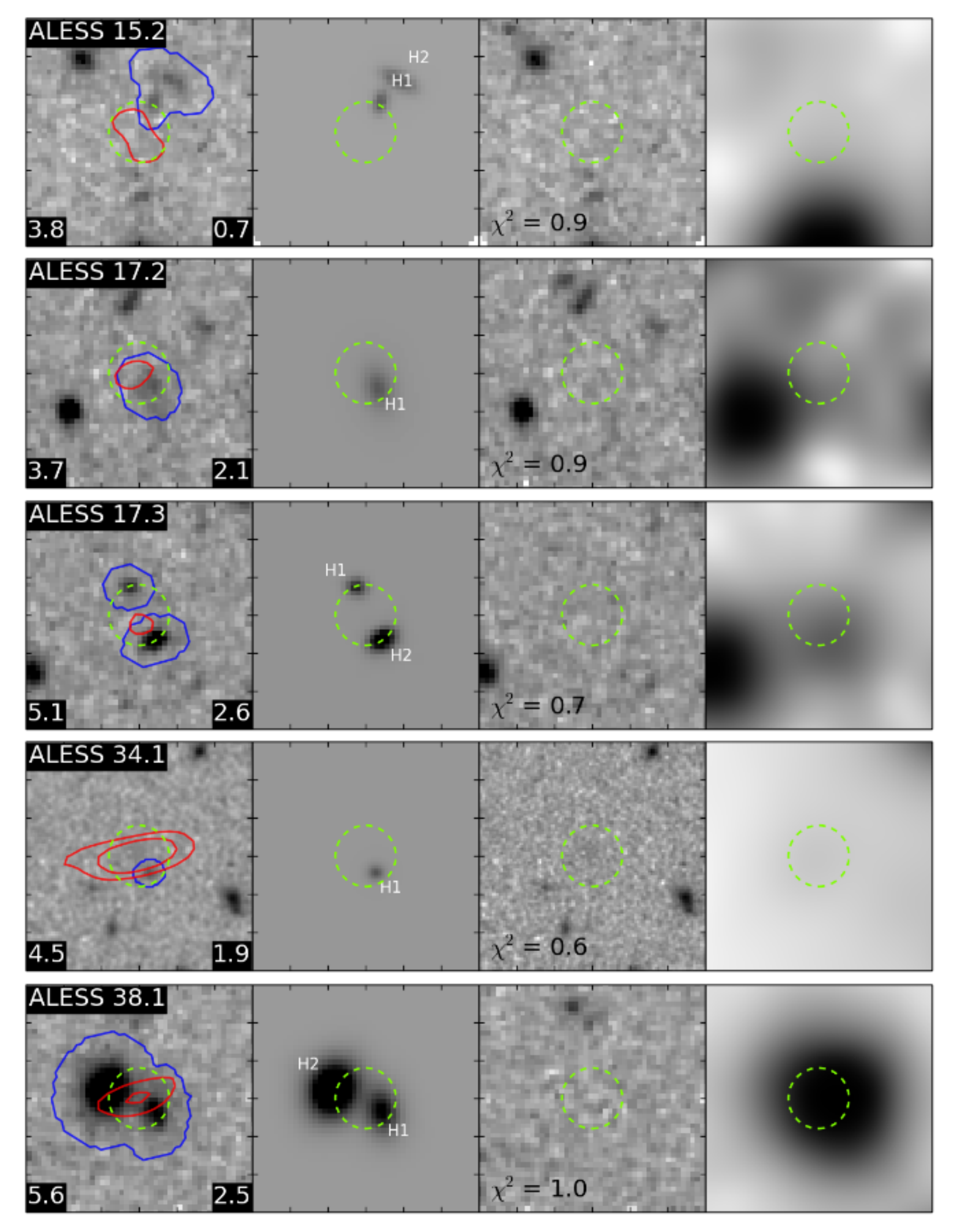}
       \caption{Same with \autoref{postage}, but on the SUPPLEMENTARY ALESS SMGs. }
\end{figure*}
\setcounter{figure}{2}
\begin{figure*}
      \includegraphics[page=2]{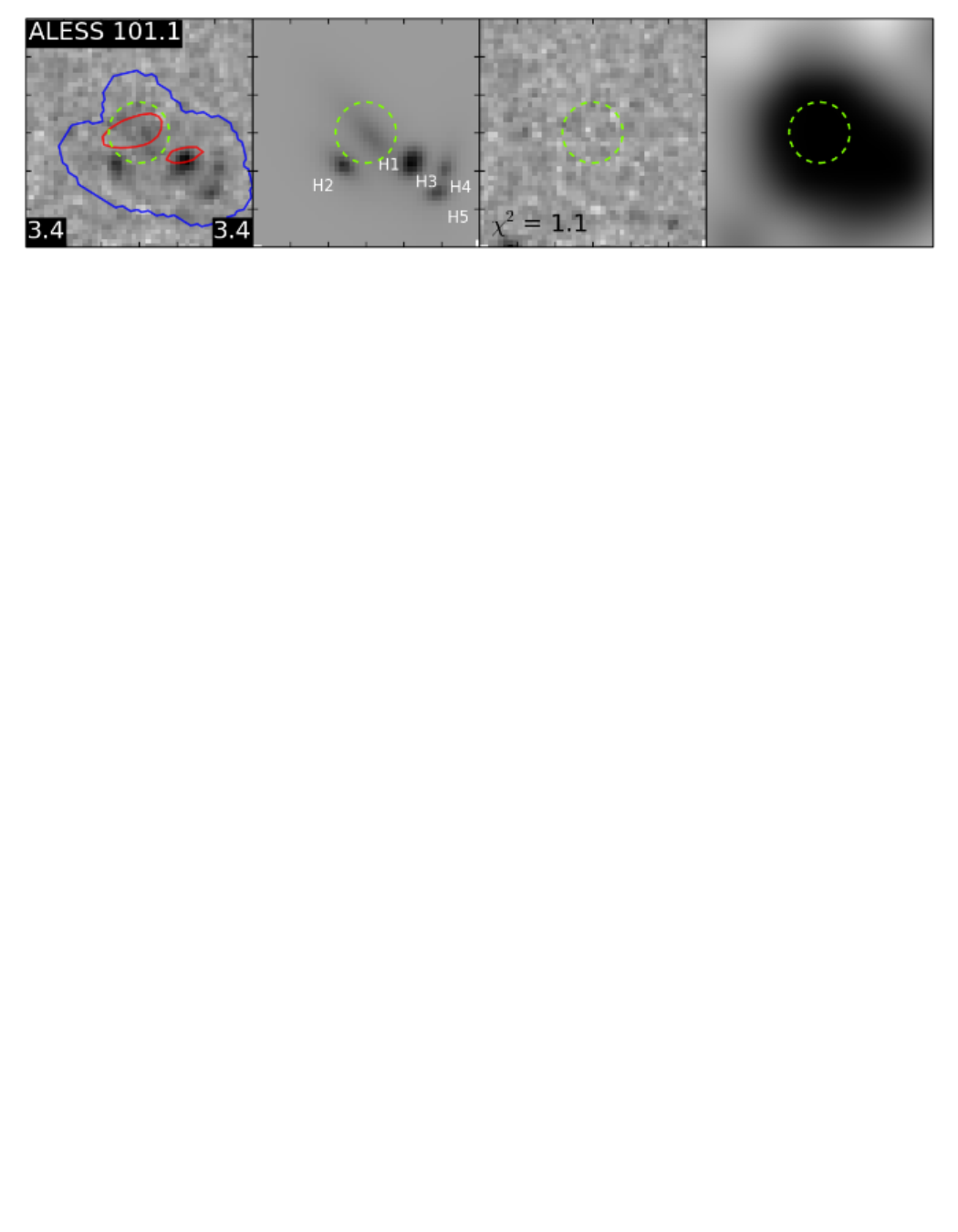}
       \caption{- {\it continued}}
\end{figure*}

\begin{figure*}
\centering
      \includegraphics[scale=1.0]{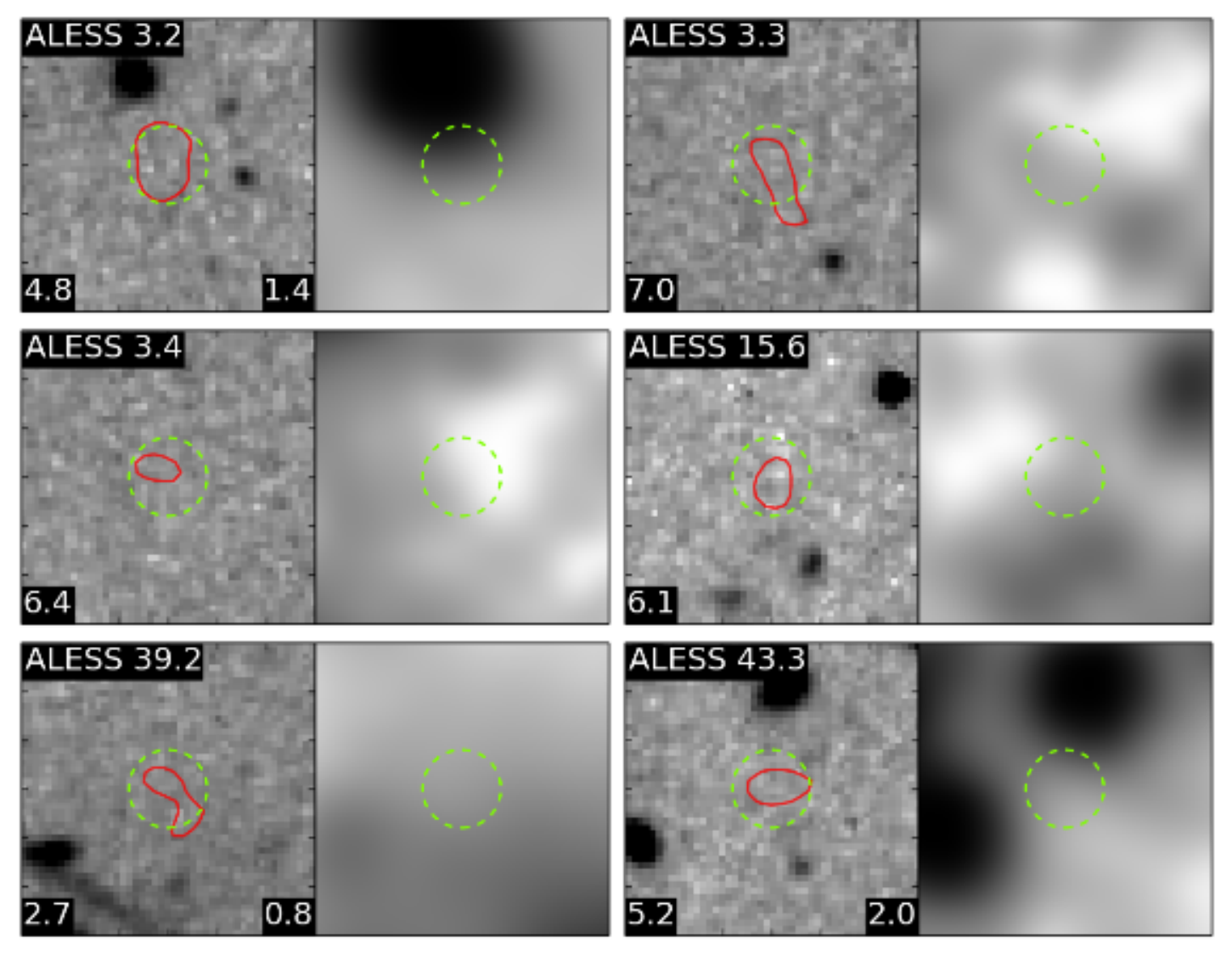}
       \caption{Same with \autoref{postage1}, but for the SUPPLEMENTARY ALESS SMGs.}
    \label{postage2}
\end{figure*}

\end{document}